\documentclass[aps,twocolumn,prl,showpacs,floatfix,altaffilletter]{revtex4-1}
\usepackage{graphicx}
\usepackage{bm}
\usepackage{amsmath}
\usepackage{sansmath}
\usepackage{amsfonts}
\usepackage{amssymb}
\usepackage{xcolor}
\usepackage{xspace}
\usepackage{xr}
\usepackage{natbib}

\begin{document}

\title{Quasiparticle Breakdown and Spin Hamiltonian of the\\
Frustrated Quantum Pyrochlore Yb$_2$Ti$_2$O$_7$ in Magnetic Field}

\author{J.~D.\ Thompson,$^{1}$ P.~A.\ McClarty,$^{2,3}$ D.\
Prabhakaran,$^{1}$ I.\ Cabrera,$^{1}$ T.\ Guidi,$^{2}$ and R.\
Coldea$^{1}$}

\affiliation{$^{1}$Clarendon Laboratory, University of Oxford,
Parks Road, Oxford OX1 3PU, United Kingdom}

\affiliation{$^{2}$ISIS Facility, Rutherford Appleton Laboratory,
Chilton, Didcot, OX11 0QX, United Kingdom}

\affiliation{$^{3}$Max Planck Institute for the Physics of Complex
Systems, N\"{o}thnitzer Str. 38, 01187 Dresden, Germany}

\pacs{75.10.Jm, 75.30.Ds}

% 75.10.Jm Quantized spin models, including quantum spin frustration
% 75.30.Ds Spin waves (for spin-wave resonance, see 76.50.+g)

\begin{abstract}
The frustrated pyrochlore magnet Yb$_2$Ti$_2$O$_7$ has the
remarkable property that it orders magnetically, but has no
propagating magnons over wide regions of the Brillouin zone. Here
we use inelastic neutron scattering to follow how the spectrum
evolves in cubic-axis magnetic fields. At high fields we observe
in addition to dispersive magnons also a two-magnon continuum,
which grows in intensity upon reducing the field and overlaps with
the one-magnon states at intermediate fields leading to strong
renormalization of the dispersion relations, and magnon decays.
Using heat capacity measurements we find that the low and high
field regions are smoothly connected with no sharp phase
transition, with the spin gap increasing monotonically in field.
Through fits to an extensive data set we re-evaluate the spin
Hamiltonian finding dominant quantum exchange terms, which we
propose are responsible for the anomalously strong fluctuations
and quasiparticle breakdown effects observed at low fields.
\end{abstract}

\maketitle

%%%%%%%%%%%%%%%%%%%%%%%%%%%%%%%%%%%%%%%%%%%%%%%%%%%%%%%%%%%%%
%%%%%%%%%%%%%%%%%%%%%%%%%%%%%%%%%%%%%%%%%%%%%%%%%%%%%%%%%%%%%
%% INTRODUCTION
%%%%%%%%%%%%%%%%%%%%%%%%%%%%%%%%%%%%%%%%%%%%%%%%%%%%%%%%%%%%%
%%%%%%%%%%%%%%%%%%%%%%%%%%%%%%%%%%%%%%%%%%%%%%%%%%%%%%%%%%%%%
The lattice of corner-shared tetrahedra realized in cubic
A$_2$B$_2$O$_7$ pyrochlores and AB$_2$O$_4$ spinels, is a
canonical lattice to explore correlated magnetism in the presence
of strong geometric frustration effects. In the strongly
spin-orbit coupled rare earth pyrochlores, experiment has
uncovered materials offering a tremendously rich spectrum of
magnetic behavior. Notable examples include classical spin ice
physics as in the rare-earth titanates (Ho/Dy)$_2$Ti$_2$O$_7$
where Ising antiferromagnetism leads to an emergent classical
electrostatics at low temperatures \cite{Castelnovo2012} and
``order-by-disorder'' in XY antiferromagnets where thermal and
quantum fluctuations lift a large frustration-induced degeneracy
resulting in unconventional magnetic order as in Er$_2$Ti$_2$O$_7$
\cite{Savary2012,Zhitomirsky2012,Savary2012b,Ross2014}. Currently,
much of the interest in this field concentrates on a handful of
materials that seem to fall outside a semiclassical understanding
of these systems. The pyrochlore Yb$_2$Ti$_2$O$_7$
\cite{Blote,Yasui2003,Bonville2004,Gardner2004,Ross2009,Thompson2011a,Ross2011a,Chang2012,Ross2012a,Applegate2012a,PhysRevB.93.064406,hayre2013thermodynamic,Pan2014,Lhotel2014,Chang2014,Gaudet,Pan2015,Robert2015,Tokiwa2016,yaouanc2016novel},
where Kramers Yb$^{3+}$ ions behave as effective spin 1/2 moments,
is quite unique in its behavior: in high applied magnetic fields
dispersive magnons were observed \cite{Ross2011a}, which are
apparently replaced by a broad continuum of scattering at zero
field \cite{PhysRevB.93.064406} despite the presence of
ferromagnetic order. This exotic behavior is not yet understood.
To make progress one would like to know i) how the broad
scattering continuum in zero field originates from quantum
fluctuations, whether those fluctuations are also present at high
field and, if so, how they manifest themselves, ii) how the sharp
magnons ``disappear'' over a wide range of the Brillouin zone as
the field is lowered. Here we experimentally answer those
questions by studying the behavior in a magnetic field applied
along the cubic [001] direction, which has not been explored in
detail before and which, we will show, allows for a transparent
interpretation of the phase diagram and evolution of the spectrum
in a magnetic field. The experiment also allows us to re-visit the
parametrization of the magnetic exchange, which is a critical
ingredient for any future theoretical understanding.

The magnetic ground state of Yb$_2$Ti$_2$O$_7$ has ferromagnetic
polarization along one of the cubic axes and moments canted
towards the local $\langle 111\rangle$ axes
\cite{Yasui2003,Chang2012,PhysRevB.93.064406,yaouanc2016novel}.
The transition to this magnetic order is observed in the range
0.2-0.26~K depending on precise synthesis conditions, and is
absent altogether in some samples \cite{Blote,Ross2012a}. Here we
report inelastic neutron scattering (INS) and heat capacity
studies on large single crystals with a sharp magnetic transition
near $T_{\rm N}=0.214(2)$~K, similar in behavior to that observed
in the ``best'' crystals \cite{Chang2012,Arpino2017}, so we
believe the magnetic properties are representative of the
high-purity limit. We find clear evidence that quantum
fluctuations are present even at the highest fields probed (9~T)
and become progressively stronger upon lowering field. Using an
extensive data set on the high-field magnon dispersions for two
distinct field orientations combined with magnetization data we
re-evaluate the spin Hamiltonian with freely-refined
nearest-neighbor exchange parameters and two $g$-tensor terms. We
find a spin Hamiltonian that is more ``quantum'' than suggested by
earlier studies \cite{Ross2011a}, and we propose that such
dominant quantum exchange terms may provide the mechanism to
explain the unexpectedly strong dispersion renormalization and
magnon decay effects at low fields, unique among all quantum
pyrochlore magnets studied so far.

\begin{figure*}[ht!]
\begin{center}
\includegraphics[width=\textwidth]{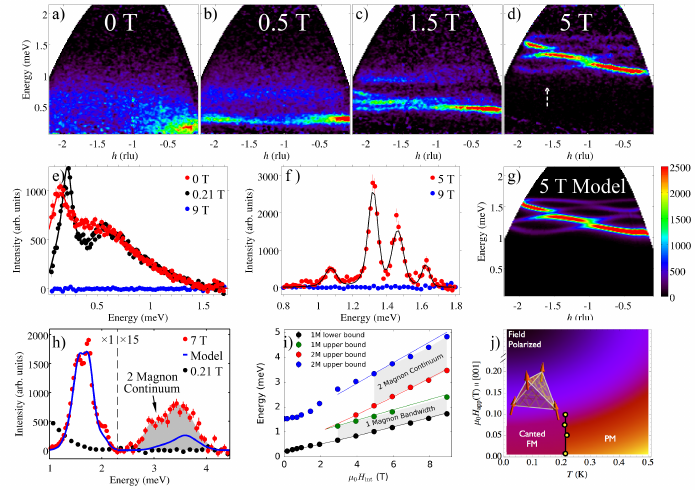}
\caption{ (a-d) Evolution of the magnetic excitation spectrum in
[001] magnetic field at 0.15~K. Horizontal axis is the in-plane
wavevector component along (100). d) The sharp modes at 5~T can be
well described in (g) by spin waves of the model Hamiltonian
(\ref{eq:Ham},\ref{eq:pars}). Vertical white arrow shows direction
along which intensity is plotted in f) for $h=[-1.8,-1.4]$, fitted
to Gaussians (solid line). h) Energy scan showing the two-magnon
scattering continuum (shaded area) at 7~T ($E_i=6.3$~meV,
$h=[-4,0]$, intensities above 2.3~meV (vertical dashed line) are
scaled $\times15$ for visibility). Solid line is the spin-wave
prediction for non-interacting magnons. All calculations include
the magnetic form factor and convolution with an estimated
resolution lineshape. i) Boundaries of the two-magnon (2M)
continuum and one-magnon (1M) extreme energies as a function of
$\mu_0H_{\rm int}$ field, lines are spin-wave calculations, which
predict overlap below 2.25~T. b-c) At 0.5 and 1.5~T the
highest-energy magnon has merged with the continuum and the three
low-energy magnons have strongly renormalized (suppressed)
dispersion. a) In zero field a broad scattering continuum
dominates, with lineshape shown in e) (red symbols, $h=[-2.1,0]$,
lines are guides to the eye. Filled blue circles illustrate
quality of the background subtraction and black symbols show that
small fields 0.21~T stabilize a sharp mode at low energies. j)
Phase diagram showing the canted ferromagnet (FM), paramagnetic
(PM) and field-polarized regions, smooth color variation indicates
crossovers and filled circles are sharp peaks in heat capacity.
Thick arrows show moment directions for a primitive tetrahedron in
the canted FM order with the magnetic field applied vertically
up.} \label{fig:1}
\end{center}
\vspace{-25pt}
\end{figure*}

The spin dynamics in a 6.3~g single crystal of Yb$_2$Ti$_2$O$_7$
were probed using the neutron spectrometer LET \cite{Bewley2011}
at the ISIS neutron source, at a temperature of $0.15$~K and in
magnetic fields up to 9~T along the [001] axis (for more details
see \cite{sm}). An estimated non-magnetic background was
subtracted from the raw neutron data and intensities were
corrected for absorption effects. Unless explicitly stated
otherwise, field values quoted throughout refer to the externally
applied fields $\mu_0 H_{\rm app}$, whereas the spin-wave
calculations are performed for the corresponding
demagnetization-corrected (internal) fields $\mu_0 H_{\rm int}$
\cite{sm}. We first discuss the results in high field where the
spectrum is dominated by strongly dispersive modes.
Fig.~\ref{fig:1}d) shows representative data at 5~T collected with
neutrons of incident energy $E_i=2.5$~meV for a fixed sample
orientation that probed the scattering at low energies for
wavevectors near ($\bar{2}00$). Sharp modes are observed, and a
representative lineshape profile is shown in Fig.~\ref{fig:1}f),
which reveals four well-defined peaks, whose positions were
extracted using Gaussian fits (solid line). By fitting similar
energy scans extracted from a volume of data collected for
multiple sample orientations an extensive data set on the
dispersion relations was obtained for many reciprocal space
directions (shown in Figs.~\ref{fig:HoraceSlices} and
\ref{fig:HoraceSlices2} panels a-d) and m-p) in the Supplemental
Material \cite{sm}).

%%%%%%%%%%%%%%%%%%%%%%%%%%%%%%%%%%%%%%%%%%%%%%%%%%%%%%%%%%%%%
%%%%%%%%%%%%%%%%%%%%%%%%%%%%%%%%%%%%%%%%%%%%%%%%%%%%%%%%%%%%%
%% ONE-MAGNON RESULTS AND ANALYSIS
%%%%%%%%%%%%%%%%%%%%%%%%%%%%%%%%%%%%%%%%%%%%%%%%%%%%%%%%%%%%%
%%%%%%%%%%%%%%%%%%%%%%%%%%%%%%%%%%%%%%%%%%%%%%%%%%%%%%%%%%%%%
The observed sharp modes are physically attributed to magnons
originating on the four sublattices of the pyrochlore structure,
which can be regarded as an FCC Bravais lattice of corner-shared
tetrahedra. Following \cite{Ross2011a} we compare the observed
dispersion relations to spin-wave modes of a nearest-neighbor
Hamiltonian with four symmetry-allowed, anisotropic exchange terms
and two $g$-tensor terms ($g_{\parallel}$ and $g_{\perp}$, along
and transverse to the local three-fold $\langle 111 \rangle$ axes)
to describe the Zeeman interaction. The exchange Hamiltonian reads
\cite{Ross2011a}
\begin{eqnarray}
{\cal H}_{\rm Exchange} & = & \sum_{\langle ij \rangle} \big\{
J_{zz} \mathsf{S}^z_i  \mathsf{S}^z_j
- J_{\pm} \left( \mathsf{S}^+_i  \mathsf{S}^-_j + \mathsf{S}^-_i  \mathsf{S}^+_j \right)  \nonumber  \\
&+& J_{\pm\pm} \left( \gamma_{ij} \mathsf{S}^+_i  \mathsf{S}^+_j + \gamma_{ij}^{*} \mathsf{S}^-_i  \mathsf{S}^-_j \right)
\nonumber \\
&+& J_{z\pm} \left[ \mathsf{S}^z_i \left( \zeta_{ij}
\mathsf{S}^+_j + \zeta_{ij}^{*} \mathsf{S}^-_i \right) + \left( i
\leftrightarrow j  \right) \right] \big\}, \label{eq:Ham}
\end{eqnarray}
where the sum extends over all nearest-neighbor $ij$ pairs counted
once ($\gamma$ and $\zeta$ are geometric terms given in
\cite{sm}). $\mathsf{S}$ denotes a spin-1/2 operator with
components expressed in a local frame where the $z$-axes point
along the local [111] direction. Following \cite{Ross2011a} we
calculate the spin-wave spectrum at high fields after finding the
mean-field ground state numerically. Varying any of the six
parameters separately
($J_{zz},J_{\pm},J_{\pm\pm},J_{z\pm},g_{\parallel},g_{\perp}$)
modifies the equilibrium spin orientation, this affects the
spin-wave expansion of the Hamiltonian and in turn the dispersion
relations, with the consequence that the parameters are all
strongly coupled making it challenging to obtain the values
independently in an unbiased way. Strongly coupled parameters
generically occur when the spin Hamiltonian does not have
rotational symmetry around the field axis, as is the case here
\cite{note}. To provide enough constraints we have extracted over
550 dispersion points combined from our data in field along
$[001]$ and previous data in field along $[\bar{1}10]$ from
\cite{Ross2011a} (the two data sets provide complementary
projections of the $g$-tensor and exchange matrices) and also
magnetization data near saturation (this further constrains the
absolute $g$-values). Using this extended data set a free fit to
the six parameters converged to a unique solution (for details see
\cite{sm})
\begin{eqnarray}
J_{zz}   = & ~0.026(3)~{\rm meV}, &J_{\pm\pm}  =  0.048(2)~{\rm meV}, \nonumber \\
J_{\pm}  = & ~ 0.074(2)~{\rm meV}, &J_{z\pm}  =  -0.159(2)~{\rm meV}, \nonumber \\
g_\parallel  = & \!\!\!\!\!\! 2.14(3), &~~g_\perp =
4.17(2).\label{eq:pars}
\end{eqnarray}
We find one dominant coupling, the ``transverse'' exchange
$J_{z\pm}$, all other exchanges are much smaller. As a consistency
check of the applicability of the spin-wave approximation we have
verified that the dispersions at 5 and 9~T are described by the
same parameters. This confirms that 5~T is a sufficiently high
field that dispersion renormalization effects beyond linear spin
wave order are negligible. The above Hamiltonian provides an
excellent description of all the data available (compare
Figs.~\ref{fig:1}d and g), Figs.~\ref{fig:HoraceSlices} and
\ref{fig:HoraceSlices2} panels a-d) with e-h) \cite{sm}. The
earlier exchange parameters proposed in \cite{Ross2011a}, deduced
assuming a larger $g$-tensor anisotropy, did not fit the [001]
field data well [compare Fig.~\ref{fig:HoraceSlices}a-d) with
i-l)]. The revised parameters fit well the dispersions data for
{\it both} field directions, and furthermore reproduce THz data on
the zone-centre magnon energies at high field [see
Fig.~\ref{fig:THz}] and the most recent estimate of the $g$-tensor
anisotropy $g_{\perp}/g_{\parallel}$ deduced from crystal-field
studies \cite{Gaudet}; the exchange parameters are also consistent
with a recent parameterization of the zero-field quasielastic
diffuse scattering pattern at higher temperature
(0.4~K)\cite{Robert2015}. A semi-classical analysis of this
Hamiltonian puts the system in a canted ferromagnetic phase, as
seen experimentally. However, the system is located very close in
parameter space to a phase boundary with an ordered XY phase
\cite{sm}. This fact may prove to be of significance
\cite{jaubert2015multiphase} in understanding the anomalously
large fluctuation effects at low field (discussed below).

%%%%%%%%%%%%%%%%%%%%%%%%%%%%%%%%%%%%%%%%%%%%%%%%%%%%%%%%%%%%%
%%%%%%%%%%%%%%%%%%%%%%%%%%%%%%%%%%%%%%%%%%%%%%%%%%%%%%%%%%%%%
%% TWO-MAGNON RESULTS AND ANALYSIS
%%%%%%%%%%%%%%%%%%%%%%%%%%%%%%%%%%%%%%%%%%%%%%%%%%%%%%%%%%%%%
%%%%%%%%%%%%%%%%%%%%%%%%%%%%%%%%%%%%%%%%%%%%%%%%%%%%%%%%%%%%%
The INS data in high field contains, in addition to sharp
one-magnon modes, also a weak scattering continuum in an energy
range corresponding to twice the one-magnon energies. This is
illustrated in Fig.~\ref{fig:1}h) at 7~T. The strong signal in the
range 1.3-2.2 meV is due to one-magnon excitations. Note the broad
signal in the range 2.75-4 meV, not present at low field (filled
circles, 0.21~T). The magnetic character of this continuum
scattering is confirmed by a strong field dependence, its energy
boundaries move in magnetic field at twice the rate compared to
the extreme energies of one-magnon states [see Fig.~\ref{fig:1}i)]
and its integrated intensity increases strongly upon decreasing
field [see Fig.~\ref{fig:ContinuumField}]. Those features are
characteristic for two-magnon excitations. If the Hamiltonian does
not have rotational invariance around the field as is the case
here, then zero-point quantum fluctuations are present at all
fields and reduce the ordered moment from its maximum value and
magnetization saturation is reached only asymptotically (as shown
in Fig.~\ref{fig:Magnetization}). In the presence of such
zero-point fluctuations neutrons can also scatter by
simultaneously creating two magnons that share the energy and
momentum transfer, thus appearing as a continuum contribution in
the INS. Upon lowering the field zero-point quantum fluctuations
are expected to grow, the magnetization to decrease and the
two-magnon scattering intensity to increase, as indeed observed.
Having established the physical origin of the continuum scattering
we note that its intensity is underestimated compared to
one-magnon states and there is more intensity at the lower
boundary than predicted by a non-interacting spin-wave model
(solid line in Fig.~\ref{fig:1}h), implying that magnon-magnon
interactions are important for describing the lineshapes
quantitatively.

Upon decreasing field the one-magnon energies decrease linearly
and the two-magnon continuum boundaries decrease at twice that
rate, see Fig.~\ref{fig:1}i). Qualitative changes in the spectrum
occur when the highest-energy magnon branch overlaps with the
continuum, expected to occur near 2.25~T. Fig.~\ref{fig:1}c) shows
data well below this field at 1.5~T, the highest-energy magnon can
no longer be distinguished from the continuum (suggesting strong
one$\rightarrow$two-magnon decay processes) and the dispersion
bandwidth of the lower three modes is strongly renormalized
(suppressed), suggesting strong interactions between one and
two-magnon states even when overlap does not occur. Upon further
lowering the field to 0.5~T, Fig.~\ref{fig:1}b), the three sharp
modes appear squeezed into a single, gapped, almost dispersionless
branch followed by strong continuum scattering at higher energies
At 0.21~T a clear gapped sharp mode can still be observed near
0.22~meV [see Fig.~\ref{fig:1}e) black circles], and at zero field
(red symbols) there is a relatively broad maximum near 0.15~meV
and a continuum lineshape extending up to $1.5$~meV. Such dramatic
quasi-particle breakdown effects over a large part of the
Brillouin zone are very unusual in three-dimensional ordered
magnets and demonstrate anomalously strong quantum fluctuations.

\begin{figure}[htbp]
\begin{center}
\includegraphics[width=\columnwidth]{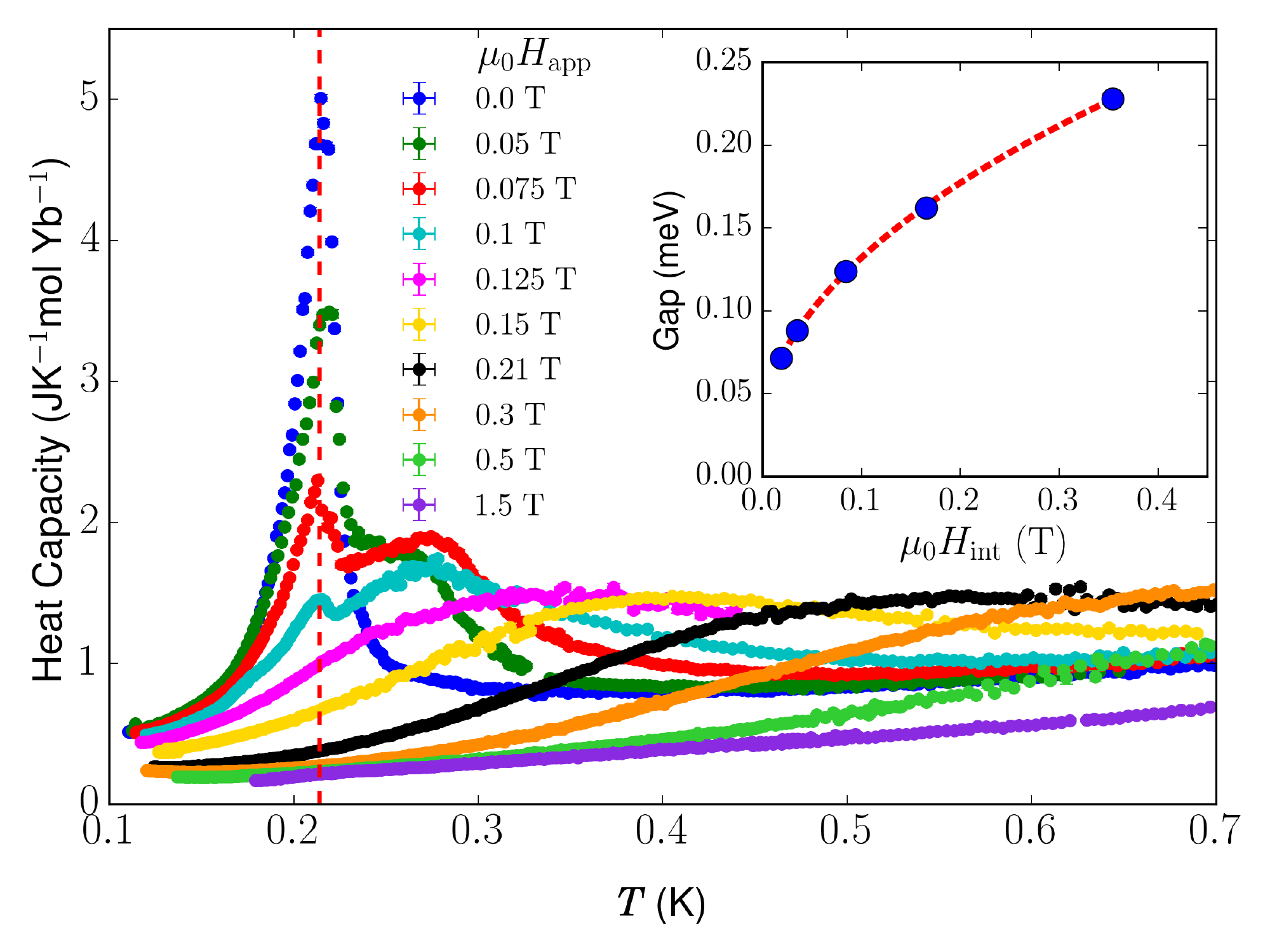}
\caption{Heat capacity as a function of temperature, $C(T)$, at
various fields $\parallel$ [001], vertical dashed line at $T_{\rm
N}$ is a guide to the eye. (inset) Gap extracted from the heat
capacity data as described in the text, dashed line is guide to
the eye.} \label{fig:2}
\end{center}
\vspace{-10pt}
\end{figure}

To obtain complementary information on the lowest-energy magnetic
excitations and evolution of the spin gap in magnetic field we
have performed AC heat capacity measurements down to 0.1~K and
fields up to 1.5~T $\parallel$ [001] on a 9.7~mg rectangular
single crystal (for details see \cite{sm}). In zero field, a sharp
anomaly is observed near 0.214(2)~K [see Fig.~\ref{fig:2} top
trace, blue symbols]. This anomaly is attributed \cite{Chang2012}
to the transition to canted ferromagnetic order with spontaneous
ferromagnetic polarization along a cubic axis. We observe that in
a very small applied field of 0.05~T (green trace) the anomaly is
significantly reduced and a broad hump appears at higher
temperatures near 0.25~K. Increasing the field to 0.1~T (cyan
trace) the anomaly is almost completely suppressed and has
disappeared at 0.125~T, where only a broad Schottky feature is
observed, which moves to higher temperature upon increasing the
field. Plotting the location of the sharp anomaly on a
field-temperature phase diagram in Fig.~\ref{fig:1}j) shows that
the phase transition from paramagnet to the low-temperature canted
ferromagnet exists only over a very small field range, above which
no phase transition boundary exists. This can be naturally
understood as follows. The transition to the canted ferromagnet
spontaneously picks the direction of the ferromagnetic
polarization along one of the cubic axes (6 domains) and the
canting of moments is uniquely determined for each site in
relation to the orientation of the local 3-fold axes. In an
external magnetic field along [001], the direction of the
ferromagnetic polarization is picked by the field $--$ the
paramagnet and the ordered phase have the same symmetry $--$ so
there is no longer a need for a phase transition. If the
transition in zero field were continuous, one would expect it to
occur only at strictly zero field. If it were first order, as is
believed to be the case here, one would expect it to survive for a
small, but finite field range, which is fully consistent with our
data. The phase diagram in Fig.~\ref{fig:1}j) shows that in [001]
field the canted ferromagnet and the high-field-polarized state
are continuously connected, without encountering a phase
transition. This is further supported by the field dependence of
the heat capacity data. The strong suppression of heat capacity at
low temperatures and presence of a Schottky anomaly that moves to
higher temperatures upon increasing field are characteristic
signatures of a spin gap that increases monotonically upon
increasing field. To parameterize this behavior we have compared
the rising part of the $C(T)$ data (up to the broad peak) to the
form expected for a two-level system (for details of fits see
\cite{sm}). The extracted gap is plotted in
Fig.~\ref{fig:2}(inset) and shows a rapid increase in field.
%increases more rapidly than for non-interacting moments
%$g_{[001]\mu_{\rm B}}\mu_0H_{\rm int}$ (lower dashed line),
%indicating the importance of interactions and/or fluctuations.
The monotonic gap increase is consistent with no phase boundary
between the canted ferromagnet and the high-field-polarized state
in Fig.~\ref{fig:1}j).

%% Conclusions
To summarize, we have reported high-resolution INS measurements of
the spin dynamics as a function of applied magnetic field in the
frustrated pyrochlore Yb$_2$Ti$_2$O$_7$. We have observed direct
evidence for coherent quantum fluctuations manifested in a
field-dependent two-magnon scattering continuum at high fields,
and strong magnon decay and dispersion renormalization effects at
low fields. Through fits to an extensive data set of the
high-field magnon dispersions and magnetization data we have
re-evaluated the spin Hamiltonian finding dominant quantum
exchange terms, which we propose are responsible for the
anomalously strong quantum fluctuation effects observed at low
field. It may be the case that those effects may be understood as
being related to the close proximity of the material to the
semi-classical phase boundary between canted ferromagnet and an
order-by-disorder antiferromagnet, or potentially a quantum spin
liquid phase. We propose that the experimental strategy employed
here of probing quasi-particle breakdown in fields comparable to
the exchange strength will be a useful experimental tool in future
studies of quantum spin liquid candidates. Our results emphasize
the need for theoretical efforts to understand the quantum phase
diagram of effective spin-1/2 pyrochlore Hamiltonians away from
the well-understood semiclassical limit.

JDT was supported by the University of Oxford Clarendon Fund
Scholarship and NSERC of Canada. PAM acknowledges support from an
STFC Keeley-Rutherford fellowship. Work in Oxford was partially
supported by EPSRC (UK) through grants No. EP/H014934/1 and
EP/M020517/1.

\bibliography{library}

\section{Supplemental Material}
\makeatletter
\renewcommand{\thefigure}{S\@arabic\c@figure}
\renewcommand{\theequation}{S\@arabic\c@equation}
\renewcommand{\thetable}{S\@arabic\c@table}
\makeatother
\setcounter{figure}{0}
\setcounter{equation}{0}

Here we provide additional technical details on 1) the anisotropic
spin exchange Hamiltonian on the pyrochlore lattice, 2) the linear
spin-wave formalism to derive the magnon dispersion relations at
high fields, 3) time-of-flight inelastic neutron scattering (INS)
experiments to probe the spin dynamics, 4) the fitting procedure
to extract the Hamiltonian parameters from high-field one-magnon
dispersion relations and magnetization, and comparison with THz
data, 5) semiclassical calculations for the fitted Hamiltonian:
mean-field ordering temperature, Curie-Weiss temperature,
mean-field phase diagram, proximity of Yb$_2$Ti$_2$O$_7$ to the
phase boundary between canted ferromagnetic and antiferromagnetic
orders, 6) observation of a two-magnon scattering continuum at
high field and comparison with spin-wave predictions, 7) magnon
dispersion renormalization and decay effects when one- and
two-magnon phase spaces overlap, 8) ac heat capacity measurements
and spin gap dependence in field, 9) magnetization measurements.

\section{S1. Spin Hamiltonian}
%%%%%%%%%%%%%%%%%%%%%%%%%%%%%%%%%%%%
\subsection{Pyrochlore Lattice and Local Frame}
%%%%%%%%%%%%%%%%%%%%%%%%%%%%%%%%%%%%
\label{sec:lattice-frame}

The pyrochlore lattice may be viewed as an FCC Bravais lattice
with a tetrahedral basis. The basis is taken to be $\bm{r}_1 =
(a/8)(1,1,1)$, $\bm{r}_2 = (a/8)(1,-1,-1)$, $\bm{r}_3 =
(a/8)(-1,1,-1)$, $\bm{r}_4 = (a/8)(-1,-1,1)$ with coordinates
given in the global (cubic axes) frame, where $a$ is the cubic
lattice parameter. A general site on the pyrochlore lattice
labelled $i$ is located at position $\bm{R}_i + \bm{r}_n$, where
$\bm{R}_i$ is the FCC lattice position and $n=1$ to $4$ is the
sublattice index. A natural choice of coordinate frame on the
pyrochlore lattice has the local $z$ axis at each site along the
local $[111]$ direction, distinguished by the local oxygen
environment. In particular, we take the local right-handed frames
for the four sublattices with unit vectors as
\begin{eqnarray}
\bm{\hat{z}}_1 & = \frac{1}{\sqrt{3}} (1,1,1), ~~~ \bm{\hat{z}}_2 & = \frac{1}{\sqrt{3}} (1,-1,-1) \nonumber\\
\bm{\hat{z}}_3 & = \frac{1}{\sqrt{3}} (-1,1,-1),~~~ \bm{\hat{z}}_4
& = \frac{1}{\sqrt{3}} (-1,-1,1) \label{eq:z}
\end{eqnarray}
and
\begin{eqnarray}
\bm{\hat{x}}_1 & = \frac{1}{\sqrt{6}} (-2,1,1), ~~~ \bm{\hat{x}}_2 & = \frac{1}{\sqrt{6}} (-2,-1,-1) \nonumber\\
\bm{\hat{x}}_3 & = \frac{1}{\sqrt{6}} (2,1,-1), ~~~ \bm{\hat{x}}_4
& = \frac{1}{\sqrt{6}} (2,-1,1). \label{eq:x}
\end{eqnarray}

We define a set of rotation matrices to transform vector
components from the global frame to those in the local frame of
each sublattice as
$R_{n}^{x\beta}=\bm{\hat{x}}_n\cdot\bm{\hat{e}}_\beta$,
$R_{n}^{y\beta}=\bm{\hat{y}}_n\cdot\bm{\hat{e}}_\beta$,
$R_{n}^{z\beta}=\bm{\hat{z}}_n\cdot\bm{\hat{e}}_\beta$. Here
$\bm{\hat{e}}_\beta$ stands for a unit vector axis of the global
frame, $\beta=x,y$ or $z$.

%%%%%%%%%%%%%%%%%%%%%%%%%%%%%%%%%%%%
\subsection{Anisotropic Exchange Couplings}
The most general, symmetry-allowed exchange matrix between
nearest-neighbor spins on the pyrochlore lattice is uniquely
defined by four independent terms
\cite{McClarty2009,Thompson2011a,Ross2011a,Onoda2011a}. Following
Ref.~\cite{Ross2011a} the exchange Hamiltonian expressed in the
local frames has the form in (\ref{eq:Ham})
%\begin{eqnarray} {\cal H}_{\rm Exchange}
%& = & \sum_{\langle ij \rangle} \big\{
%    J_{zz} \mathsf{S}^z_i  \mathsf{S}^z_j
%    - J_{\pm} \left( \mathsf{S}^+_i  \mathsf{S}^-_j + \mathsf{S}^-_i  \mathsf{S}^+_j \right)  \nonumber  \\
%    &  + & J_{\pm\pm} \left( \gamma_{ij} \mathsf{S}^+_i  \mathsf{S}^+_j +
%          \gamma_{ij}^{*} \mathsf{S}^-_i  \mathsf{S}^-_j \right)   \nonumber \\
%    &  + & J_{z\pm} \left[ \mathsf{S}^z_i \left( \zeta_{ij}   \mathsf{S}^+_j + \zeta_{ij}^{*} \mathsf{S}^-_i \right)
%    + \left( i \leftrightarrow j \right) \right] \big\},
%\label{eq:Hexchange}
%\end{eqnarray}
where
\begin{equation}
\zeta = \left(  \begin{array}{cccc} 0  &  -1 & e^{i\pi/3} &  e^{-i\pi/3} \\
-1  &  0 & e^{-i\pi/3} &  e^{i\pi/3} \\ e^{i\pi/3}  &  e^{-i\pi/3} & 0 &  -1 \\
e^{-i\pi/3}  &  e^{i\pi/3} & -1 &  0   \end{array} \right)
\nonumber
\end{equation}
and $\gamma=-\zeta^{*}$. In a more abbreviated notation, ${\cal
H}_{\rm Exchange}  = \sum_{\langle ij \rangle}
\mathcal{J}_{ij}^{\alpha\beta} \mathsf{S}^\alpha_i
\mathsf{S}^\beta_j$ where throughout we use the notation
convention that repeated axes indices are summed over. The spin
operators are understood to operate on the effective spin one-half
doublet at each magnetic site that characterizes the ground state
of the crystal field split $J=7/2$ multiplet of the Kramers
Yb$^{3+}$ ions. We may neglect the influence of excited crystal
field levels because the crystal field gap to the first excited
level far exceeds the exchange coupling strengths \cite{Gaudet}.
We neglect the long-range dipolar interaction because it is small
compared to the exchange. The quality of our fits (presented
below) is such that we do not require the inclusion of exchange
couplings beyond nearest neighbor.

The total Hamiltonian including the Zeeman coupling to an external
magnetic field is
\begin{equation}
{\cal H} = {\cal H}_{\rm Exchange} + {\cal H}_{\rm Zeeman}
\label{eq:Hamiltonian}
\end{equation}
with
\begin{equation}
{\cal H}_{\rm Zeeman} = - \mu_{\rm B} \mu_0 H_{\rm int}^{\mu}
\sum_{i} g^{\mu\nu}_{n} \mathsf{S}_{i}^{\nu}, \label{eq:zeeman}
\end{equation}
where the magnetic field components $\mu_{0}H_{\rm int}^{\mu}$ are
given in the global frame and $g_n^{\mu\nu}$ is the $g$-tensor for
sublattice $n$.

%%%%%%%%%%%%%%%%%%%%%%%%%%%%%%%%%%%%
\section{S2. Spin Wave Theory}
\label{sec:SWT}
%%%%%%%%%%%%%%%%%%%%%%%%%%%%%%%%%%%%
% Linear Spin Waves
% Dynamical Structure Factor

Starting from the spin Hamiltonian with effective spin one-half
moments in (\ref{eq:Hamiltonian}), we first compute the ground
state in a $[001]$ applied magnetic field within mean field
theory, assuming that the magnetic structure can always be
described using the primitive tetrahedral structural cell
(magnetic propagation vector $\bm{q}=\bm{0}$), as is the case
semiclassically at both zero and very large fields. The
orientation of the effective spin one-half moments within the
ground state ${\bm{\hat{u}}}_n$ allows us to specify a local
quantization frame on each sublattice, where
$\bm{\hat{\tilde{z}}}_n={\bm{\hat{u}}}_n$,
$\bm{\hat{\tilde{x}}}_n={\bm{\hat{u}}}_n \times [1,1,1]/\parallel
{\bm{\hat{u}}}_n \times [1,1,1] \parallel$ and
$\bm{\hat{\tilde{y}}}_n=\bm{\hat{\tilde{z}}}_n \times
\bm{\hat{\tilde{x}}}_n$. In addition to the rotation matrices
$R_n^{\alpha\beta}$ defined above, which rotate from the global to
the local frame, rotations from the local frame to the
quantization frame are given by $\bar{R}_n^{\mu\alpha}$ where
$\mu$ is the quantization frame index, $\alpha$ is the local frame
index and $n$ is the sublattice label.

In the local quantization frame, we write the spin operators in
terms of Holstein-Primakoff (HP) bosons and expand around the
classical ground state in powers of $1/S$.
\begin{align*}
\tilde{\mathsf{S}}^{z}_i & = S - a^\dagger_i a_i \\
\tilde{\mathsf{S}}^+_i & = \tilde{\mathsf{S}}^x_i+ i \tilde{\mathsf{S}}^y_i  = \sqrt{  2S - a^\dagger_i a_i } a_i \approx \sqrt{2S} a_i  \\
\tilde{\mathsf{S}}^-_i & = \tilde{\mathsf{S}}^x_i- i
\tilde{\mathsf{S}}^y_i  =  a_i^\dagger  \sqrt{  2S - a^\dagger_i
a_i }\approx \sqrt{2S} a_i^\dagger.
\end{align*}
The commutation relations satisfied by the boson operators are
\[  \left[  a_i , a_j^\dagger \right] = \delta_{ij}.   \]
Other commutators vanish.

The leading term in this expansion is the classical Hamiltonian
\[   \mathcal{H}_0 =  S(S+1) \sum_{\langle ij \rangle} \mathcal{J}_{ij}^{\alpha\beta} u_n^\alpha u_{n'}^\beta,   \]
where the subscripts $n$ and $n'$ are the sublattice indices of
the $i$ and $j$ sites, respectively. Here the term of order $S^2$
is the mean-field exchange energy and the term of order $S$ comes
from symmetrizing the exchange Hamiltonian quadratic in the boson
operators.

We write the boson operators in Fourier space. The interaction
matrix is then
\[  \mathcal{J}_{nn'\bm{k}}^{\alpha\beta} = \frac{1}{N} \sum_{ij} \mathcal{J}_{ij}^{\alpha\beta} \exp\left[ i \bm{k}\cdot (\bm{R}_i - \bm{R}_j +\bm{r}_n -\bm{r}_{n'})
\right], \] where $N$ is the number of FCC lattice sites. In the
quantization frame, the interaction matrix becomes
$\tilde{\mathcal{J}}_{nn'\bm{k}}^{\mu\nu}=\bar{R}_n^{\mu\alpha}\bar{R}_{n'}^{\nu\beta}\mathcal{J}_{nn'\bm{k}}^{\alpha\beta}$.

The terms linear in the bosons vanish in the ground state computed
from the minimum semiclassical energy and fluctuations around the
ground state may be computed from the quadratic Hamiltonian
$\mathcal{H}_2$. We introduce operators
\begin{align}
\bm{\sigma}^x_{\bm{k}} \equiv  \sqrt{ \frac{S}{2} }
\left(  \bm{a}_{\bm{k}}  + \bm{a}^\dagger_{-\bm{k}}  \right) \nonumber  \\
\bm{\sigma}^y_{\bm{k}} \equiv i \sqrt{\frac{S}{2} } \left(
\bm{a}_{\bm{k}} - \bm{a}^\dagger_{-\bm{k}} \right) \nonumber
 \end{align}
with
$\bm{a}_{\bm{k}}=(a_{1\bm{k}},a_{2\bm{k}},a_{3\bm{k}},a_{4\bm{k}})$
so that
 \begin{align}
\mathcal{H}_2 & = \sum_{\bm{k}} \left( \begin{array}{cc}
(\bm{\sigma}^x_{\bm{k}})^\dagger &
(\bm{\sigma}^y_{\bm{k}})^\dagger \end{array} \right) \left(
\begin{array}{cc}  \bm{P}^x_{\bm{k}}
& \bm{T}_{\bm{k}} \\
\bm{T}^\star_{\bm{k}}  & \bm{P}^y_{\bm{k}}
\end{array}  \right)
\left( \begin{array}{c} \bm{\sigma}_{\bm{k}}^x \\
\bm{\sigma}_{\bm{k}}^y \end{array} \right),
\nonumber \\
& \equiv \sum_{\bm{k}} \left( \begin{array}{cc}
(\bm{\sigma}^x_{\bm{k}})^\dagger &
(\bm{\sigma}^y_{\bm{k}})^\dagger \end{array} \right)
\boldsymbol{H} \left( \begin{array}{c}
\bm{\sigma}_{\bm{k}}^x \\
\bm{\sigma}_{\bm{k}}^y \end{array} \right), \label{eq:H2}
\end{align}
where the sum runs over all wavevectors $\bm{k}$ in the first
Brillouin zone of the FCC lattice and where
\begin{align*}
P^x_{nn'\bm{k}} & = \mathcal{R}_{n} \delta_{nn'} + \tilde{\mathcal{J}}_{nn'\bm{k}}^{xx} + \frac{\mu_{\rm B}}{2} \mu_0 H_{\rm int}^{\alpha}R_n^{\beta\alpha} g^{\beta} \bar{R}_n^{z\beta} \delta_{nn'}  \\
P^y_{nn'\bm{k}} & = \mathcal{R}_{n} \delta_{nn'} + \tilde{\mathcal{J}}_{nn'\bm{k}}^{yy} + \frac{\mu_{\rm B}}{2} \mu_0 H_{\rm int}^{\alpha}R_n^{\beta\alpha} g^{\beta} \bar{R}_n^{z\beta} \delta_{nn'}  \\
T_{nn'\bm{k}} & = \tilde{\mathcal{J}}_{nn'\bm{k}}^{xy}
\end{align*}
and $\mathcal{R}_n = -\sum_{n'}
\tilde{\mathcal{J}}_{nn',\bm{k}=\bm{0}}^{zz} $. Since the
$g$-tensor is diagonal in the local frame, with the form $g={\rm
diag}(g_{\perp},g_{\perp},g_{\parallel})$, we write $g$ with a
single index denoting the diagonal element.

The diagonalization of the quadratic Hamiltonian $\mathcal{H}_2$
in (\ref{eq:H2}) to find the magnon wavefunctions and energies
must preserve the boson commutation relations, which take the form
\[  \left[ \bm{\sigma}_{\bm{k}}^\alpha,\left( \bm{\sigma}_{\bm{k}}^\beta  \right)^\dagger \right]=
S \bm{\eta}^{\alpha\beta}  \] with
\[  \bm{\eta} \equiv
\left(  \begin{array}{cc} 0 & -i \bm{I}  \\
i\bm{I}  & 0 \end{array}  \right),   \] where $\bm{I}$ is the
$4\times 4$ identity matrix. Then the spin wave energies
$\omega_{m\bm{k}}$ for modes $m=1$ to 4 are the positive
semidefinite set of eigenvalues of the matrix $2\bm{\eta H}$ and
the right eigenvectors $\bm{v}_{m\bm{k}}$ of $\bm{\eta H}$
preserve the commutation relation among the $\bm{\sigma}$
operators provided that $\bm{v}^\dagger \bm{\eta} \bm{v}=\bm{g}$,
where $\bm{g} = {\rm diag}(1,1,1,1,-1,-1,-1,-1)$.

The neutron scattering intensity is proportional to
\[ \left\vert F(\bm{Q}) \right\vert^{2}
\sum_{\alpha,\beta} \left( \delta_{\alpha\beta}- \hat{Q}_{\alpha}
\hat{Q}_{\beta} \right) S^{\alpha\beta}\left( \bm{Q},\omega
\right),   \] where $\bm{Q}=\bm{k}_i-\bm{k}_f$ is the wavevector
transfer, $\bm{k}_i$ ($\bm{k}_f$) is the incident (final) neutron
wavevector, $\omega$ is the neutron energy transfer and
$F(\bm{Q})$ is the magnetic form factor of Yb$^{3+}$ ions, assumed
spherically symmetric. The indices $\alpha$ and $\beta$ are
components of the moments in the global frame. We evaluate the
inelastic part of the scattering function at zero temperature
\begin{align}
& S^{\alpha\beta}_{\rm inelas}\left( \bm{Q},\omega \right)=
\nonumber \\ & ~~~~\sum_{n,n'} \sum_E  \langle 0 \vert
\mathsf{J}^{\alpha}_n(-\bm{Q}) \vert E \rangle \langle E \vert
\mathsf{J}^{\beta}_{n'}(\bm{Q}) \vert 0  \rangle \delta\left(
\omega - \omega_E\right), \label{eq:Sinelas}
\end{align}
where the second sum is over all excited states $|E\rangle$ of
energy $\omega_E$ above the ground state $|0\rangle$. The physical
moment $\mathsf{J}^{\alpha}_n$ is related to the effective spin
one-half moment components in the quantization frame
$\tilde{\mathsf{S}}_n^\gamma$ through $\mathsf{J}^{\alpha}_n =
\Gamma_n^{\alpha\gamma} \tilde{\mathsf{S}}^{\gamma}_n$ where we
have defined $\Gamma_n^{\alpha\gamma}\equiv R_n^{\beta\alpha}
g^{\beta} \bar{R}_n^{\gamma\beta}$.

One-magnon excited states are accessed via spin fluctuations with
transverse polarization ($\mu,\nu=x$ or $y$) in the quantization
frame, and the corresponding inelastic scattering function is
\begin{align}
S^{\alpha\beta}_{\rm inelas, 1M}&\left( \bm{Q},\omega \right) =
   \frac{S}{2}  \sum_{n,n'} \sum_{\mu\nu =x,y} \Gamma_n^{\alpha\mu}
\Gamma_{n'}^{\beta\nu}  \times \nonumber
\\ & \sum_{m}   v^{\mu \dagger}_{mn,-\bm{k} } v^{\nu}_{m n' \bm{k}}
e^{-i\bm{\tau}\cdot \bm{r}_{nn'} }
\delta(\omega-\omega_{m\bm{k}}),
\end{align}
where $\bm{r}_{nn'}=\bm{r}_n-\bm{r}_{n'}$ and the eigenvectors are
written as $v^{\mu}_{mn\bm{k}}$, indexed by the transverse spin
deviation component $\mu=x,y$. Here $\bm{k}$ is the wavevector
transfer reduced to the first Brillouin zone, i.e.
$\bm{Q}=\bm{\tau}+\bm{k}$ where $\bm{\tau}$ is closest reciprocal
lattice vector to $\bm{Q}$.

Two-magnon excited states are accessed via fluctuations polarized
longitudinally in the quantization frame. To see this we consider
the longitudinal scattering function for the effective spin in the
quantization frame, which has an analogous form to
(\ref{eq:Sinelas})
\begin{equation}
S_{{\rm eff},nn'}^{zz}(\bm{Q},\omega) \equiv \sum_E \langle 0
\vert \tilde{ \mathsf{S}}^{z}_n (-\bm{Q}) \vert E \rangle \langle
E \vert  \tilde{ \mathsf{S}}^{z}_{n'} (\bm{Q}) \vert 0 \rangle
\delta(\omega - \omega_E), \label{eq:longitudinal}
\end{equation}
where
\begin{align} \tilde{ \mathsf{S}}^{z}_n (\bm{Q}) & =
\sqrt{N}S \delta(\bm{k}) e^{-i\bm{\tau}\cdot \bm{r}_n}  \nonumber \\
& - \frac{1}{\sqrt{N}}\sum_{\bm{q},\bm{q'}}
\delta(\bm{k}-\bm{q}+\bm{q'}) a_{n\bm{q}}^\dagger a_{n\bm{q'}}
e^{-i\bm{\tau}\cdot \bm{r}_n} \label{eq:szz}
\end{align}
is the component of the effective spin along the quantization
direction. The first term leads to a contribution of to order
$S^2$ in the structure factor (\ref{eq:longitudinal}), which is
the elastic, Bragg scattering in the ordered phase. The second
term in (\ref{eq:szz}) leads to an inelastic contribution of order
$S^0$ in the structure factor (\ref{eq:longitudinal}), due to
magnon pair creation/annihilation.

We introduce
\begin{eqnarray}
\Lambda_{mn\boldsymbol{q}} & \equiv (v^{x }_{mn\boldsymbol{q}}
-i v^{y }_{mn\boldsymbol{q}})/2 \nonumber \\
\bar{\Lambda}_{mn\boldsymbol{q}} & \equiv (v^{x
}_{mn\boldsymbol{q}} + i v^{y }_{mn\boldsymbol{q}})/2.
\label{eq:lambda}
\end{eqnarray}
It is also convenient to introduce $\Omega_{\bm{q}\bm{q'}nmm'}
\equiv  \Lambda_{mn\bm{q}} \bar{\Lambda}_{m'n\bm{q'}} +
\bar{\Lambda}_{mn\bm{q}} \Lambda_{m'n\bm{q'}}
(1-(2-\sqrt{2})\delta_{mm'}\delta(\bm{q'}-\bm{q}))$ such that the
inelastic part of (\ref{eq:longitudinal}) becomes
\begin{align*}
S&_{{\rm eff, inelas}, nn'}^{zz}(\bm{Q},\omega)  =  \frac{1}{N}
\sum_{\bm{q},\bm{q'}} \sum_{m\geq m'}
\Omega^{\star}_{\bm{q}\bm{q'}nmm'} \Omega_{\bm{q}\bm{q'}n'mm'} \times \\
& ~~~~~e^{-i\bm{\tau}\cdot \bm{r}_{nn'}}
\delta(\bm{k}-\bm{q}-\bm{q'})
\delta\left(\omega-\omega_{m\bm{q}}-\omega_{m'\bm{q'}}\right).\end{align*}
The two-magnon scattering from the physical moment is given by
\[  S^{\alpha\beta}_{\rm inelas,2M}\left( \bm{Q},\omega \right) = \sum_{n,n'}  \Gamma_n^{\alpha z} \Gamma_{n'}^{\beta z} S_{{\rm eff, inelas},nn'}^{zz}(\bm{Q},\omega), \]
which we have evaluated numerically using a Monte Carlo
integration over the Brillouin zone in order to compare with the
experimental data.

%%%%%%%%%%%%%%%%%%%%%%%%%%%%%%%%%%%%%%%%%%%%%%%%%%%%%%%%%%%%%%%%%%%%%%%%
\section{S3. Inelastic Neutron Scattering Experiments}
\label{InelasticNeutrons}

\subsection{Experimental Details}
\label{neutronexperimentdets}

\begin{figure*}[htbp!]
\begin{center}
\includegraphics[width=\textwidth]{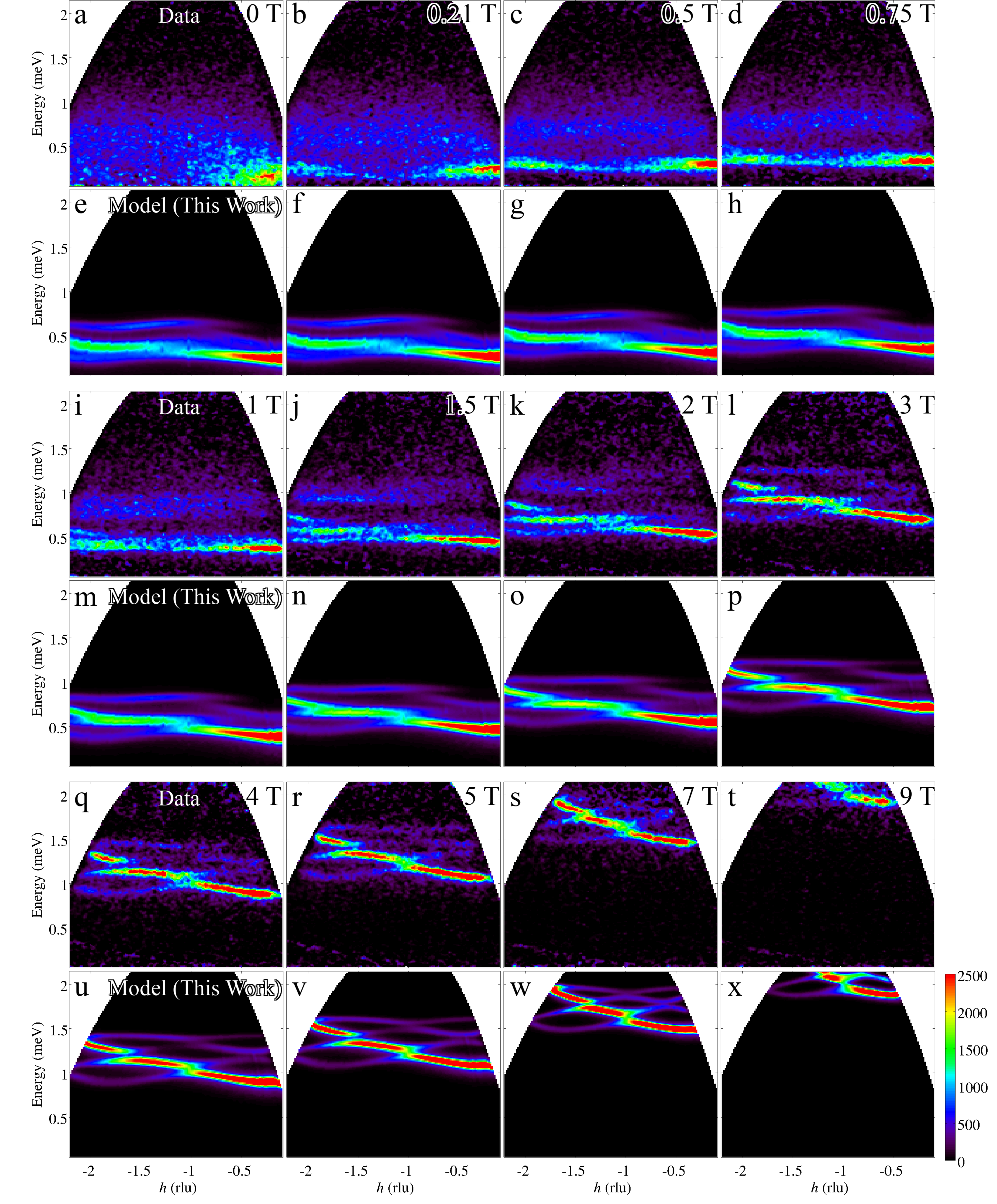}
\caption{Overview of the spin dynamics as a function of field
along $[001]$. The top panel and every other subsequent row of
panels shows raw data, paired with the corresponding spin-wave
calculation for the model Hamiltonian. The data was collected for
$E_i=2.5$~meV and horizontal axis shows the in-plane wavevector
component along the (100) direction (intensities are averaged for
the vertical component in the range $l=[-0.5,0.5]$). Calculations
are performed for the estimated internal fields in
Table~\ref{tab:demag_ins} and include the magnetic form factor and
convolution with an estimated resolution lineshape.}
\label{fig:FieldSlices2.5}
\end{center}
\end{figure*}

\begin{figure*}[htbp!]
\begin{center}
\includegraphics[width=\textwidth]{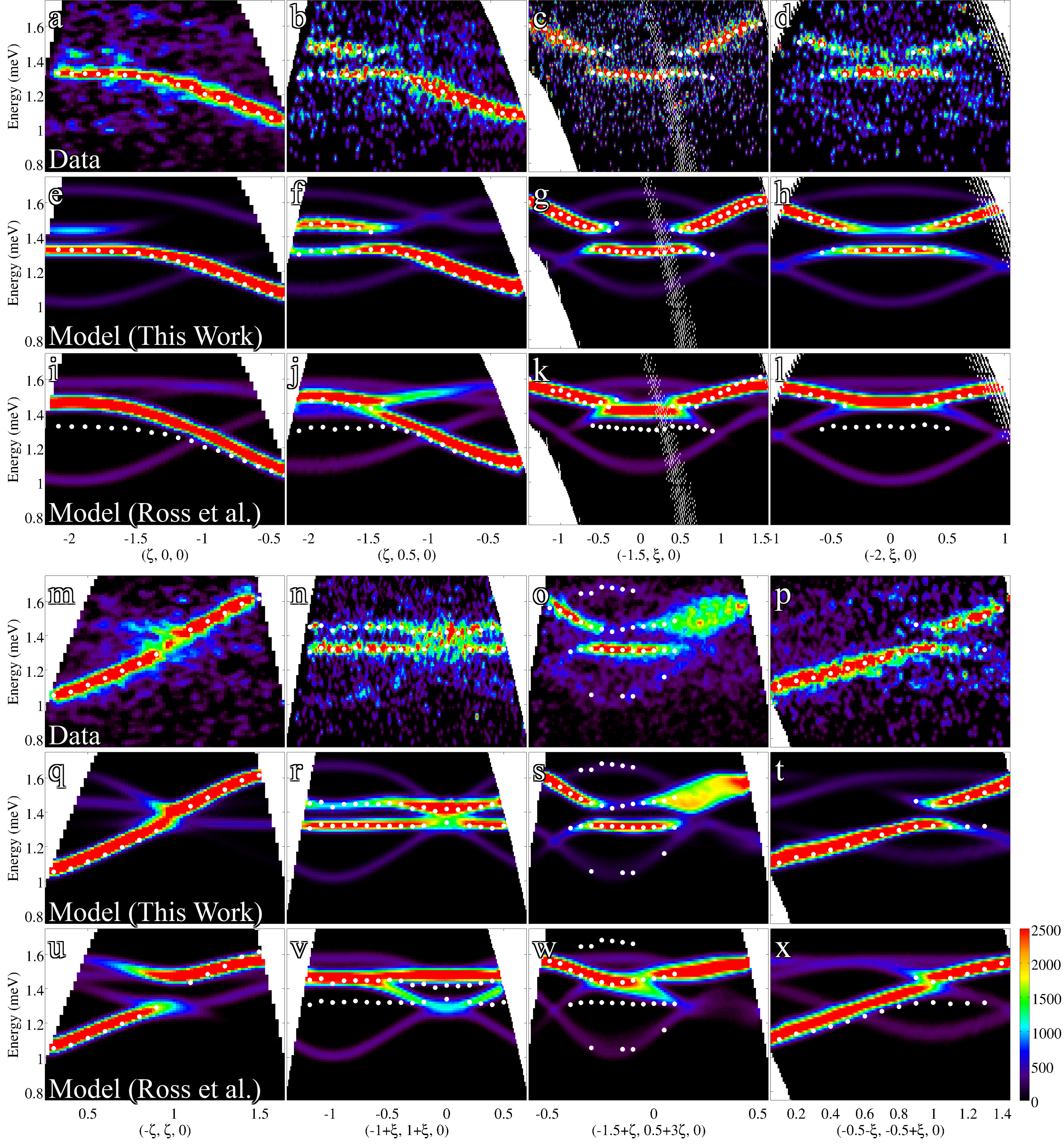}
\caption{Representative dispersion maps at 5~T $\parallel[001]$
along various directions in the ($hk0$) plane (red lines in
Fig.~\ref{fig:BZmap}). Panels (a-d,m-p) show experimental data
($E_i=2.5$~meV), the row of panels immediately below (e-h,q-t)
show the corresponding spin wave calculation for the model
Hamiltonian in (\ref{eq:pars}), whereas the subsequent row of
panels (i-l,u-x) show the calculation for the model in Ref.\
\cite{Ross2011a}. The white solid dots show the location of
experimentally-extracted dispersion points, determined by fitting
Gaussian peaks to energy scans at constant wavevector.
Calculations are performed for a demagnetization-corrected field
$\mu_0H_{\rm int}=4.93$~T and include the magnetic form factor and
convolution with an estimated energy resolution. Intensities are
averaged for transverse wavevectors in the range $l,k=[-0.2,0.2]$
for the ($h00$) direction, and similar ranges for the other
directions.} \label{fig:HoraceSlices}
\end{center}
\end{figure*}

\begin{figure*}[htbp!]
\begin{center}
\includegraphics[width=\textwidth]{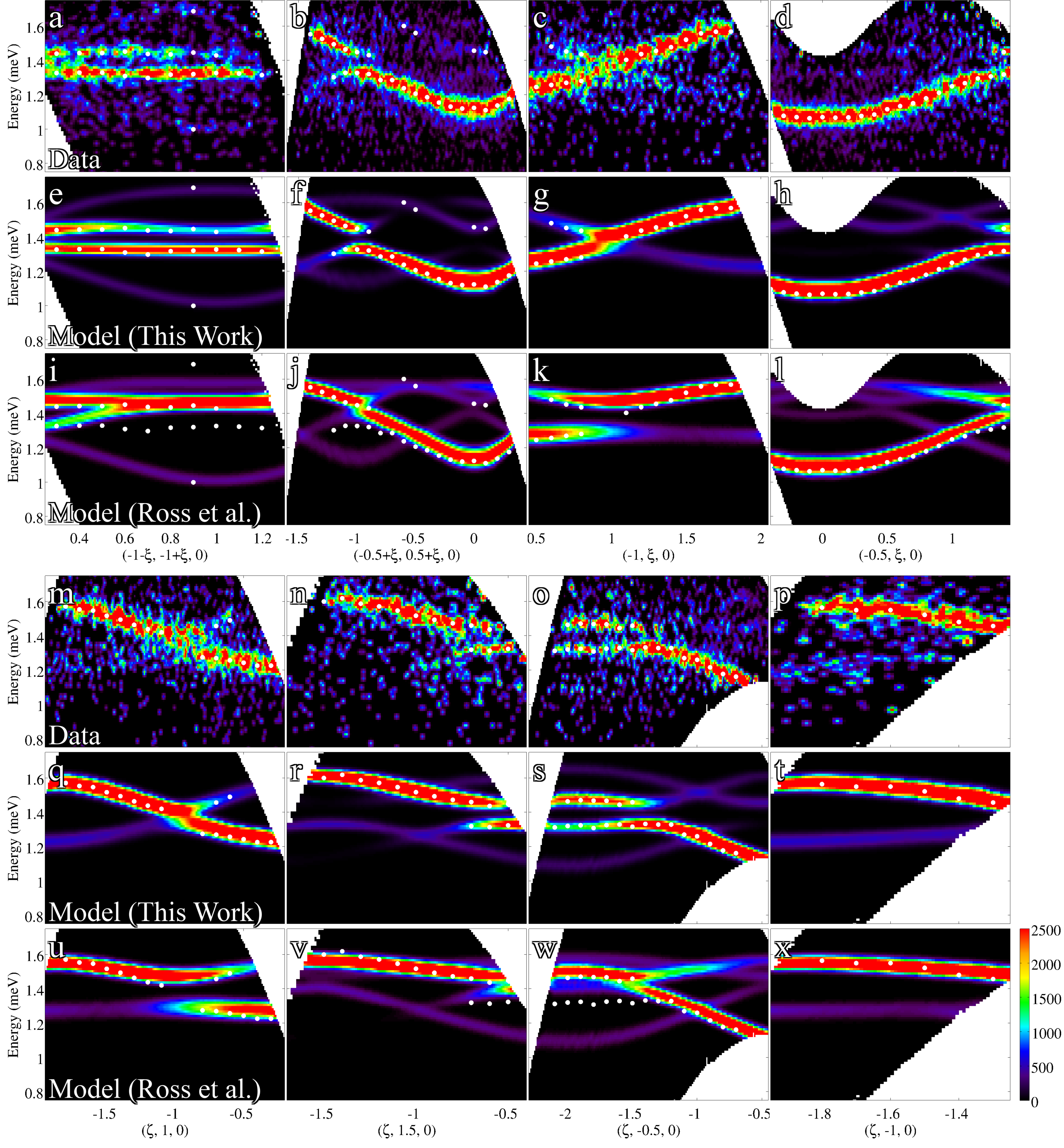}
\caption{Same as Fig.~\ref{fig:HoraceSlices}, but for other
directions in the ($hk0$) plane.}\label{fig:HoraceSlices2}
\end{center}
\end{figure*}

A single crystal of Yb$_2$Ti$_2$O$_7$ was grown as described in
Ref.~\cite{Prabhakaran2011} using a four-mirror optical
floating-zone furnace (Crystal System Inc.) in an argon rich
atmosphere with a growth rate of 1-2~mm/h, similar to the
conditions reported in Ref.~\cite{Chang2012}. Due to the argon
atmosphere, the as-grown crystal was oxygen deficient and dark in
color. In order to improve the oxygen stoichiometry, the as-grown
crystal was annealed at 1200$^{\circ}$C for 5 days under oxygen
flow atmosphere and the crystal become transparent and almost
colorless. Several pieces were cut from this larger crystal and
used for all the different measurements reported here.

\begin{figure}[tbp!]
\begin{center}
\includegraphics[width=0.8\columnwidth]{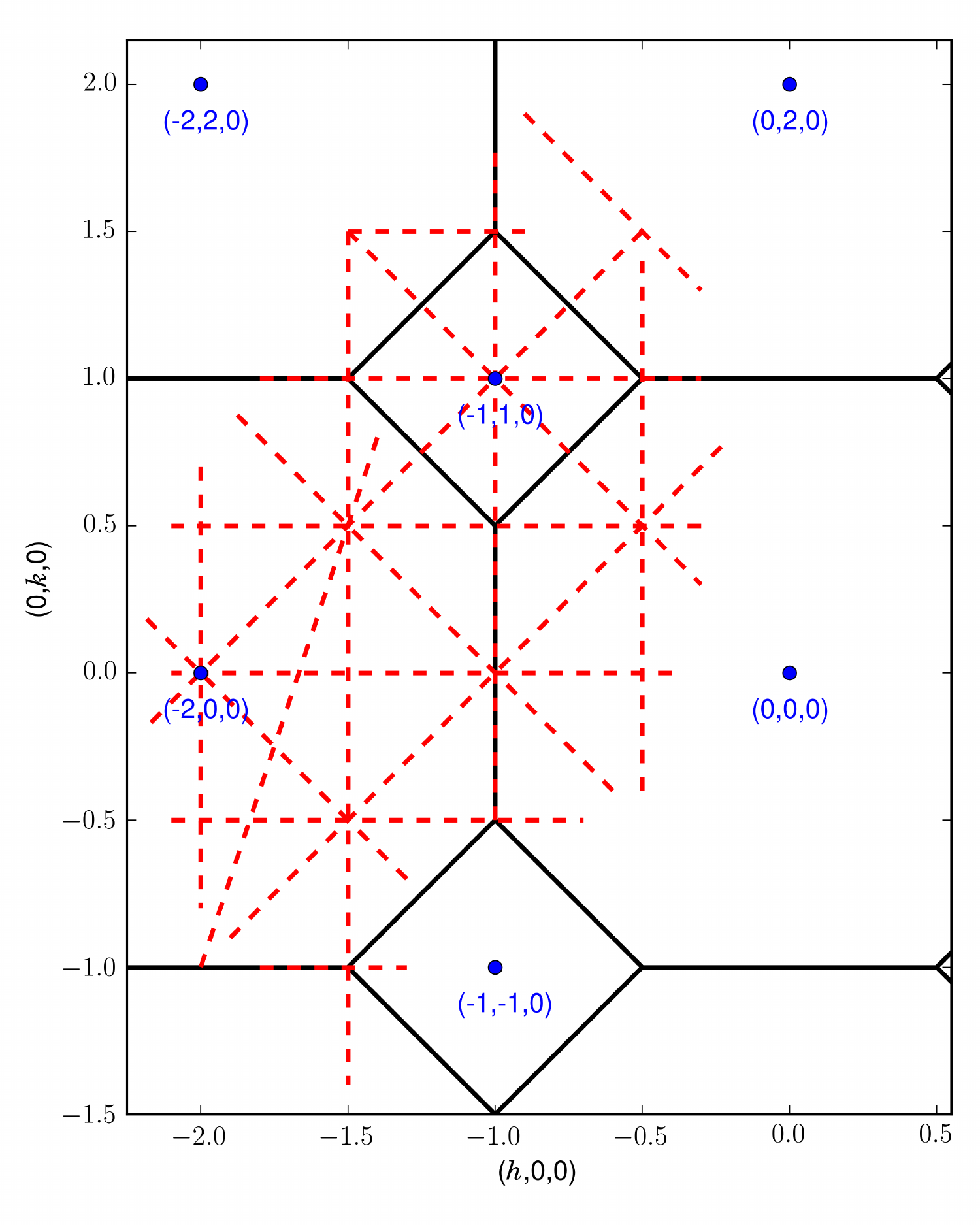}
\caption{Schematic of the ($hk0$) horizontal scattering plane
indicating where dispersion maps were extracted from Horace scans
(red lines) and plotted in Figs.~\ref{fig:HoraceSlices} and
\ref{fig:HoraceSlices2} panels a-d) and m-p). Solid black lines
show intersections with the FCC Brillouin zone boundaries.}
\label{fig:BZmap}
\end{center}
\end{figure}

The spin dynamics was probed using the direct-geometry
time-of-flight chopper spectrometer LET at the ISIS neutron source
\cite{Bewley2011} using a cylindrical-shaped $6.3$~g single
crystal of Yb$_2$Ti$_2$O$_7$ aligned in the $(hk0)$ horizontal
scattering plane. The sample was cooled using a dilution fridge
insert with a base temperature of 0.15~K where all measurements
were performed, this temperature is well below the zero-field
magnetic ordering transition temperature of 0.214(2)~K [see phase
diagram in Fig.~\ref{fig:1}j)]. Magnetic fields up to 9~T were
applied along the crystal [001] axis using a vertical cryomagnet.
Applied fields were corrected for demagnetization effects as
discussed in Sec.~S9. To avoid complexities associated with
multiple (ferro)magnetic domains the sample was cooled to base
temperature in a finite magnetic field (0.21~T) to ensure a single
magnetic domain with ferromagnetic polarization along the field.
The zero-field data was collected last, after reducing the field
to zero from finite values. The inelastic scattering was probed
for neutrons with incident energy $E_i=1.34$, $2.5$, 4 and
$6.3$~meV, with measured energy resolutions of $0.023(1)$,
$0.055(2)$, $0.094(1)$ and $0.220(5)$~meV (Full Width Half
Maximum), respectively, on the elastic line. For the majority of
the measurements LET was operated in multi-repetition mode,
allowing the mapping of the inelastic scattering with $E_i=1.34$,
2.5 and 6.3~meV simultaneously. For an overview of how the spin
dynamics evolves as a function of field, measurements were
performed up to 9~T for a fixed sample orientation that probed the
scattering at low energies for wavevectors near ($\bar{2}00$),
with a typical counting time of $2.5$~h per setting; those results
are summarized in Fig.~\ref{fig:FieldSlices2.5} (top and every
other subsequent row). To obtain an extended data set on the
wavevector-dependence of the spin dynamics in the Brillouin zone
the inelastic scattering was measured at a selection of fixed
applied fields (0, 0.21, 1.5 and 5~T) for a range of sample
rotation angles around the vertical [001] direction spanning
$90^\circ$ in steps of $1^\circ$ (Horace scan), each position
counted for approximately $7$~minutes. This gave access to a wide
range of wavevectors in the ($hk0$) plane and typical intensity
maps extracted from this data volume are shown in
Fig.~\ref{fig:HoraceSlices}a-d). We estimated the non-magnetic
energy-dependent background using Fig.~\ref{fig:1}i) as a guide to
indicate where no magnetic scattering is expected at different
fields, i.e. at high fields no magnetic signal is expected below
the one-magnon gap and in the interval between the one- and
two-magnon energy ranges (shaded regions in Fig.~\ref{fig:1}i),
whereas at low field no magnetic signal is expected at very high
energies. The estimated non-magnetic background was subtracted
from the raw intensities to obtain the purely magnetic signal,
which was afterwards corrected for neutron absorption effects
using a numerical Monte Carlo routine for a tilted cylindrical
sample (assuming an inverse velocity dependence of the neutron
absorption cross-section). Throughout this paper, wavevectors are
given as $(h,k,l)$ in reciprocal lattice units of $2\pi/a$ of the
structural cubic unit cell with lattice parameter
$a=10.026$~\AA~\cite{Gardner2004}.

\begin{figure}[htbp!]
\begin{center}
\includegraphics[width=\columnwidth]{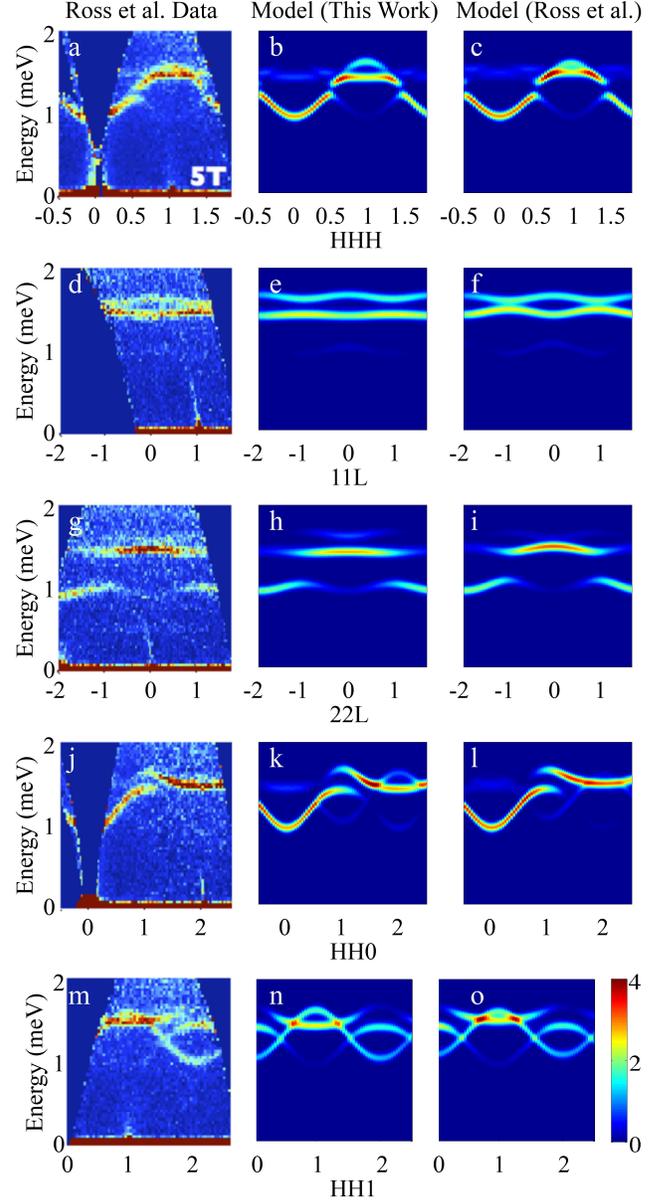}
\caption{Dispersion maps along various directions in the ($hhl$)
plane for a field of 5~T along [$\bar{1}10$] (first column is INS
data from \cite{Ross2011a}), compared with spin-wave calculations
for the refined Hamiltonian in (\ref{eq:pars})(middle column) and
the model in Ref~\cite{Ross2011a}. Both calculations include the
magnetic form factor and convolution with a finite energy
resolution of 0.09~meV (FWHM). Spin-wave calculations using the
applied field value of 5~T or an estimated
demagnetization-corrected field of $\mu_0H_{\rm int}=4.93$~T gave
essentially indistinguishable results (the latter is plotted in
the figure).} \label{fig:RossComparison}
\end{center}
\end{figure}

%%%%%%%%%%%%%%%%%%%%%%%%%%%%%%%%%%%%
\begin{figure*}[tbp!]
\begin{center}
\includegraphics[width=\textwidth]{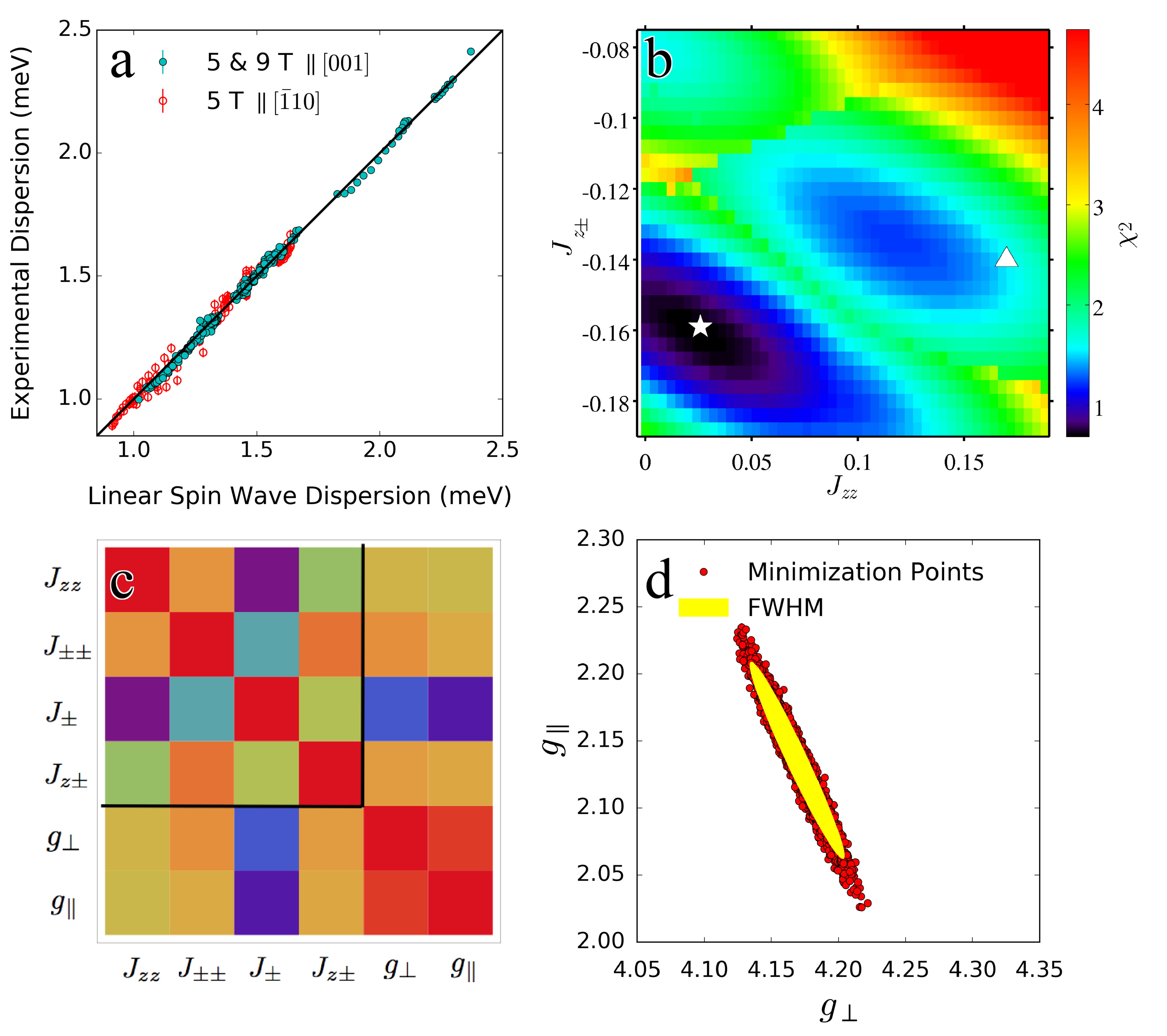}
\caption{a) Agreement between experimental and calculated
spin-wave energies for the fitted Hamiltonian in (\ref{eq:pars}).
The graph contains over 550 data points, filled cyan symbols
correspond to fields of $5$ and $9$~T $\parallel [001]$ and red
open circles to field of $5$~T $\parallel [\bar{1}10]$, the latter
from from \cite{Ross2011a}. The solid line corresponds to perfect
agreement. b) Map of the goodness of fit ($\chi^2$) as a function
of $J_{zz}$ and $J_{z\pm}$, where for each point the remaining
(four) Hamiltonian parameters are freely optimized against all
data points in a) and the magnetization constraint. The location
of the optimal parameter set (\ref{eq:pars}) is indicated by the
white star in the middle of the dark region (global minimum). The
white triangle (middle of right-hand side) corresponds to the
parameter values in Ref.~\cite{Ross2011a}. c) Absolute value of
the correlation coefficient between the six fitted parameters. d)
Visual illustration of the correlation between $g_{\parallel}$ and
$g_{\perp}$, red dots illustrate the parameter values when the
experimentally-determined spin-wave energies are randomly shifted
within their errorbars, yellow elliptical contour shows the
half-maximum contour of this distribution.} \label{fig:chisquared}
\end{center}
\end{figure*}

\subsection{High-field Spin-wave Dispersions}

Representative intensity maps for some high-symmetry directions in
the ($hk0$) plane are shown in Figs.~\ref{fig:HoraceSlices} and
\ref{fig:HoraceSlices2} panels a-d) and m-p). In those figures the
row of panels immediately below the raw data, e-h) and q-t), are
the spin-wave calculation for the fitted Hamiltonian
(\ref{eq:pars}), and demonstrate an excellent agreement with the
data along all directions probed. For comparison, the subsequent
row of panels, i-l) and u-x), show the calculation for the
parameters in Ref.~\cite{Ross2011a}, in this case systematic
differences are seen in terms of shifts of the dispersion modes
between the data and the model predictions, compare for example
Fig.~\ref{fig:HoraceSlices}a-d) with i-l), from which we conclude
that the earlier proposed model cannot account for the observed
dispersions in field along [$001$], whereas the current refined
model can account for all the dispersion data, even for field
along [$\bar{1}10$], see Fig.~\ref{fig:RossComparison}.

%%%%%%%%%%%%%%%%%%%%%%%%%%%%%%%%%%%%%%%%%%%%%%%%%%%%%%%%%%%%%%%%%%%%
\subsection{S4. Fitting Procedure to Determine the Spin Hamiltonian}
%%%%%%%%%%%%%%%%%%%%%%%%%%%%%%%%%%%%%%%%%%%%%%%%%%%%%%%%%%%%%%%%%%%%

In this section we give details of the fitting procedure used to
obtain the Hamiltonian parameters from experimentally measured
magnon dispersion relations at high fields. From the INS data in
applied fields of $5$ and $9$~T $\parallel[001]$ we extracted
energy scans [as in Fig.~\ref{fig:1}e)] and determined mode
energies $\omega_m$, where the subscript $m=1$ to 4 labels the
modes at a given ($h,k,l$) in order of increasing energy. For some
scans it was not possible to detect four distinct modes, but by
continuity with neighboring regions in reciprocal space, we were
able to determine the appropriate labels of all the visible modes.
This way we obtained a list of dispersion points
$(h,k,l,\omega,m)$. In order to provide multiple and independent
constraints on the fits, we also extracted dispersion points from
the INS data reported in \cite{Ross2011a} in a field of
$5$~T$\parallel[\bar{1}10]$. We also included in the fitting
procedure the requirement for the model to reproduce the measured
magnetization value at $\mu_0H_{\rm int}=6.86$~T $\parallel$
[001], at which field the magnetization is almost saturated (see
Sec.~S9).

By computing the magnon dispersion relations within linear spin
wave theory, we carried out a least squares minimization allowing
all six parameters of the Hamiltonian to vary independently. The
minimization was based on a simulated annealing algorithm that was
run several thousand times, initialized every time with different
random starting parameters, for an extensive sampling of the
parameter space to detect the global minimum.

As a first test, we fixed the $g$ tensor ratio
$g_{\perp}/g_{\parallel}=2.4$ as in \cite{Ross2011a} and the
minimization procedure using only the $[\bar{1}10]$ dispersion
data set converged, within error, to the parameter values given in
that paper. We also found that the $5$~T$\parallel[001]$
dispersions data alone were insufficiently constraining to
determine both $g$ factor elements. In other words, the
minimization procedure on this data set alone, without a
constraint on the $g$-factor components leads to almost degenerate
solutions with different sets of parameters. However, by taking
the $5$~T$\parallel [\bar{1}10]$ dispersion data, and the $5$ and
$9$~T $\parallel[001]$ dispersions data sets, together with the
$7$~T$\parallel [001]$ magnetization, we obtained a unique,
optimum solution, with a clear global minimum in the goodness of
fit, that can account for all the data. We emphasize that when not
all the neutron data and magnetization constraints are included in
the fitting procedure, there is a family of almost degenerate
solutions to the optimization problem.

The existence of a set of parameters that nearly satisfy all the
fitting constraints is a consequence of the strong correlation
between certain parameters. In Fig.~\ref{fig:chisquared}d), we
show the magnitude of the correlation coefficient between the six
parameters. All $6$ parameters are significantly correlated with
particularly large correlations between the two $g$ factor
parameters, which can be traced back to the constraint on the
magnetization. In the absence of the magnetization constraint, the
strongest correlation is between $J_{zz}$ and $g_{\parallel}$.

The refined Hamiltonian parameters obtained using all the above
mentioned data points are given in (\ref{eq:pars}), where the
quoted uncertainties were estimated as follows. Each dispersion
point ($h,k,l,w,m$) had an energy uncertainty $\sigma$, which
ranged between $0.01$ and $0.02$~meV. The larger error bars are
for the magnon energies determined visually from the digitized INS
intensity maps reported in Ref.~\cite{Ross2011a}. We created a
list of $10^4$ sets of mode energies, where each set is shifted
from the original by a gaussian random number times the standard
deviation of the energy of that mode. We carried out the least
square minimization to determine the optimum model parameters for
each such set of slightly shifted mode energies. The errors on the
model parameters were then extracted from the resulting
distributions of those fitted parameters.

{\it Field Misalignment} $-$ In the INS measurements the magnetic
field was slightly misaligned relative to the $[001]$
crystallographic axis by $1.0(1)^{\circ}$. We have verified that
assuming perfect alignment $\parallel[001]$ or including this
small misalignment resulted in essentially identical values for
the optimized parameters, within the uncertainties listed in
(\ref{eq:pars}).

\begin{figure}[tbp!]
\includegraphics[width=0.95\columnwidth]{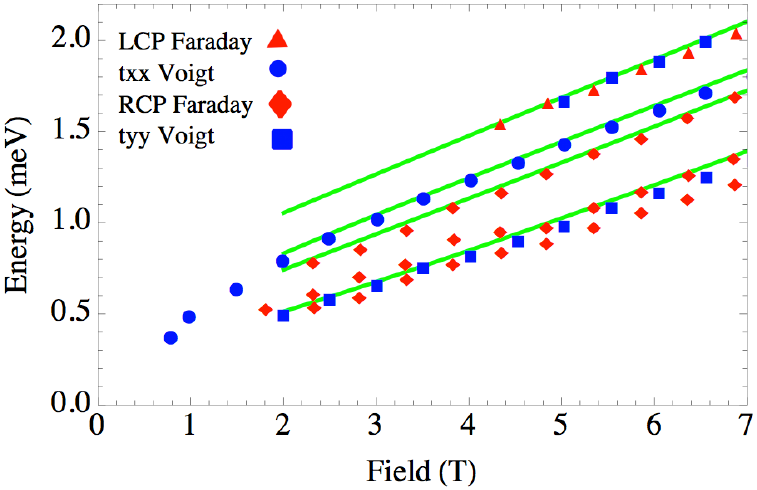}
\caption{THz spectrum as a function of magnetic field along [001]
from Ref.~\cite{Pan2014}, symbols correspond to excitation
energies observed using different experimental geometries and
polarizations as listed in the legend. Solid lines are the
calculated spin-wave energies at the zone center,
$\omega_{1-4,\bm{k}=\bm{0}}$, for the model Hamiltonian in
(\ref{eq:pars}), assuming quoted field values have negligible
demagnetization corrections.} \label{fig:THz}
\end{figure}

\subsection{Comparison with THz data}

As a further test of the determined Hamiltonian we have verified
consistency also with the recently reported THz spectrum for field
$\parallel[001]$. THz spectroscopy measures the long wavelength
($\bm{k}=\bm{0}$) excitation spectrum and in total four main
families of peaks are observed using different experimental
setups, see Fig.~\ref{fig:THz}. The energies of the observed modes
are well reproduced (solid lines) for fields above 3~T by the spin
wave energies calculated for the Hamiltonian in (\ref{eq:pars})
with no adjustable parameters. At lower fields linear spin wave
theory fails owing to the proximity of single and two-magnon
states (as discussed later in Sec.~S7).

We have also verified the applicability of the THz selection rules
to the observation of spin wave modes. THz spectroscopy measures
the complex transmittance $t(\omega)$ of initially polarized THz
radiation through a single crystal sample in the presence of a
static magnetic field. The transmittance is proportional to the
imaginary part of the susceptibility
$\chi''_{\alpha\beta}(\omega)$, which we calculate using a random
phase approximation. The Faraday geometry with the incident THz
pulse in the $z$ direction and linearly polarized in the $x$
direction and the Voigt geometry with the field in the $y$
direction, where all axes labels refer to the global (cubic axes)
frame. In the case of the Faraday geometry, Ref.~\cite{Pan2014}
concentrated on the transmittance of circularly polarized
radiation. We find that the spectrum in the Voigt geometry,
obtained from $\chi''_{xx}(\omega)$ and $\chi''_{yy}(\omega)$, is
sensitive to three magnon modes - the lowest and two highest
modes, as indeed observed (blue symbols in Fig.~\ref{fig:THz}). In
the Faraday geometry, the lowest mode should appear in the right
circularly polarized channel, from
$\chi''_{xx}(\omega)-\chi'_{xy}(\omega)$, and the highest mode in
the left circularly polarized channel,
$\chi''_{xx}(\omega)+\chi'_{xy}(\omega)$. However, the
second-to-lowest mode should not be visible using the Faraday
geometry although, apparently, this mode is visible in the data.
However, if we allow for a small field misalignment away from the
$[001]$ direction then the second-to-lowest mode exhibits a peak
in the THz spectrum, making the Faraday spectrum also consistent
with the model predictions.

%%%%%%%%%%%%%%%%%%%%%%%%%%%%%%%%%%%%
\section{S5. Semiclassical Properties of the Spin Hamiltonian}

\subsection{Quantum Fluctuations}
\label{sec:DeltaS}

The role of quantum fluctuations on the size of the ordered moment
may be computed within linear spin wave theory. The departure of
the effective spin one-half moment from its fully available value
$S=1/2$ is given by
\begin{equation}
\Delta S \equiv \frac{1}{4N} \sum_{i} \langle a_{i}^{\dagger}a_{i}
\rangle = \frac{1}{4N} \sum_{\bm{k}} \sum_{n,m}
{\Lambda}^{*}_{mn\bm{k}} {\Lambda}_{mn\bm{k}} \label{eq:DeltaS}
\end{equation}
in terms of the rotated eigenvectors $\Lambda$, defined in
(\ref{eq:lambda}). At 5~T we obtain a relatively small reduction
of 2\%, providing at least a partial consistency check to justify
the applicability of the linear spin-wave approach to parameterize
the dispersion relations and extract the Hamiltonian. We note
however that even though quantum fluctuations are small at those
fields, they are still present and are ultimately responsible for
the observation of a weak, but finite intensity two-magnon
continuum in addition to the dominant one-magnon excitations in
INS (to be discussed in detail in Sec.~S6).

%%%%%%%%%%%%%%%%%%%%%%%%%%%%%%%%%%%%
\begin{figure}[htbp!]
\includegraphics[width=0.95\columnwidth]{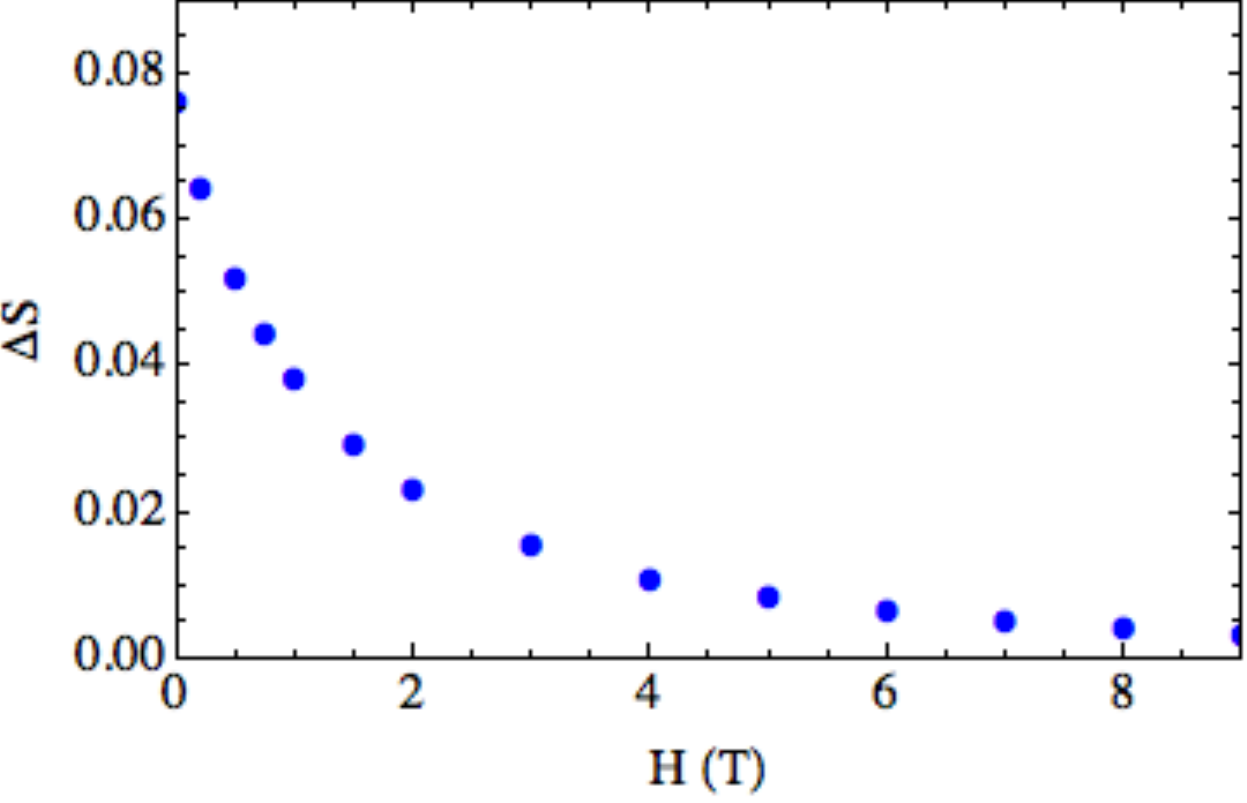}
\caption{Spin reduction $\Delta S$ as a function of applied field
calculated using linear spin-wave theory, eq.~(\ref{eq:DeltaS}).
\label{fig:DeltaS}}
\end{figure}
%%%%%%%%%%%%%%%%%%%%%%%%%%%%%%%%%%%%

\subsection{Total Moment Sum Rule}

The total moment sum rule is a single constraint on the structure
factor computed for a model with spin $S$
\begin{equation}
\frac{1}{N} \sum_{\bm{k},\alpha} \int  S_{\rm
eff}^{\alpha\alpha}(\bm{k},\omega)~d\omega=  S(S+1)
\label{eq:sum_rule}
\end{equation}
where the structure factor $S_{\rm eff}$ is computed for the
effective spin one-half moment. This is to be distinguished from
the experimental structure factor for Yb$_2$Ti$_2$O$_7$, which is
computed for the true $\mathsf{J}$ moment in the ground state
crystal field doublet.

In this section, we compile various contributions to the total
scattering sum rule for the exchange parameters determined from
experiment. We concentrate on the 9~T magnon spectrum because, of
all the measured fields, this one should be the closest match to
linear spin wave theory. In the previous section we discussed the
leading order quantum correction to the ordered spin, which is
$\Delta S =0.0030$ at 9~T. The total elastic, one- and two-magnon
contributions are listed in Table~\ref{table:TMSR}. The combined
total differs from the sum rule in (\ref{eq:sum_rule}) by less
than 0.25\%, such small violations of the sum rule are generally
expected in linear spin wave theory, higher order contributions in
$1/S$ are generally needed to renormalize the intensities to agree
with the sum rules.

\begin{table}[htb!]
\caption{Calculated contributions to total (effective spin) sum
rule at $9$ T. The elastic (Bragg) contribution is $(S-\Delta
S)^2$ where $\Delta S$ is the zero-point spin reduction in
(\ref{eq:DeltaS}). The single magnon (1M) and two magnon (2M)
contributions are obtained from numerical integration of the
transverse and longitudinal (effective spin) dynamical
correlations, respectively.} \centering
\begin{tabular}{c c c c c}
\hline
& Elastic & 1M & 2M & Total \\ [0.5ex] % inserts table %heading
\hline
Intensity & 0.241 & 0.50244 & 0.0051 & 0.74854 \\
\hline
\end{tabular}
\label{table:TMSR}
\end{table}

%%%%%%%%%%%%%%%%%%%%%%%%%%%%%%%%%%%%
\begin{figure}[htbp!]
\includegraphics[width=0.95\columnwidth]{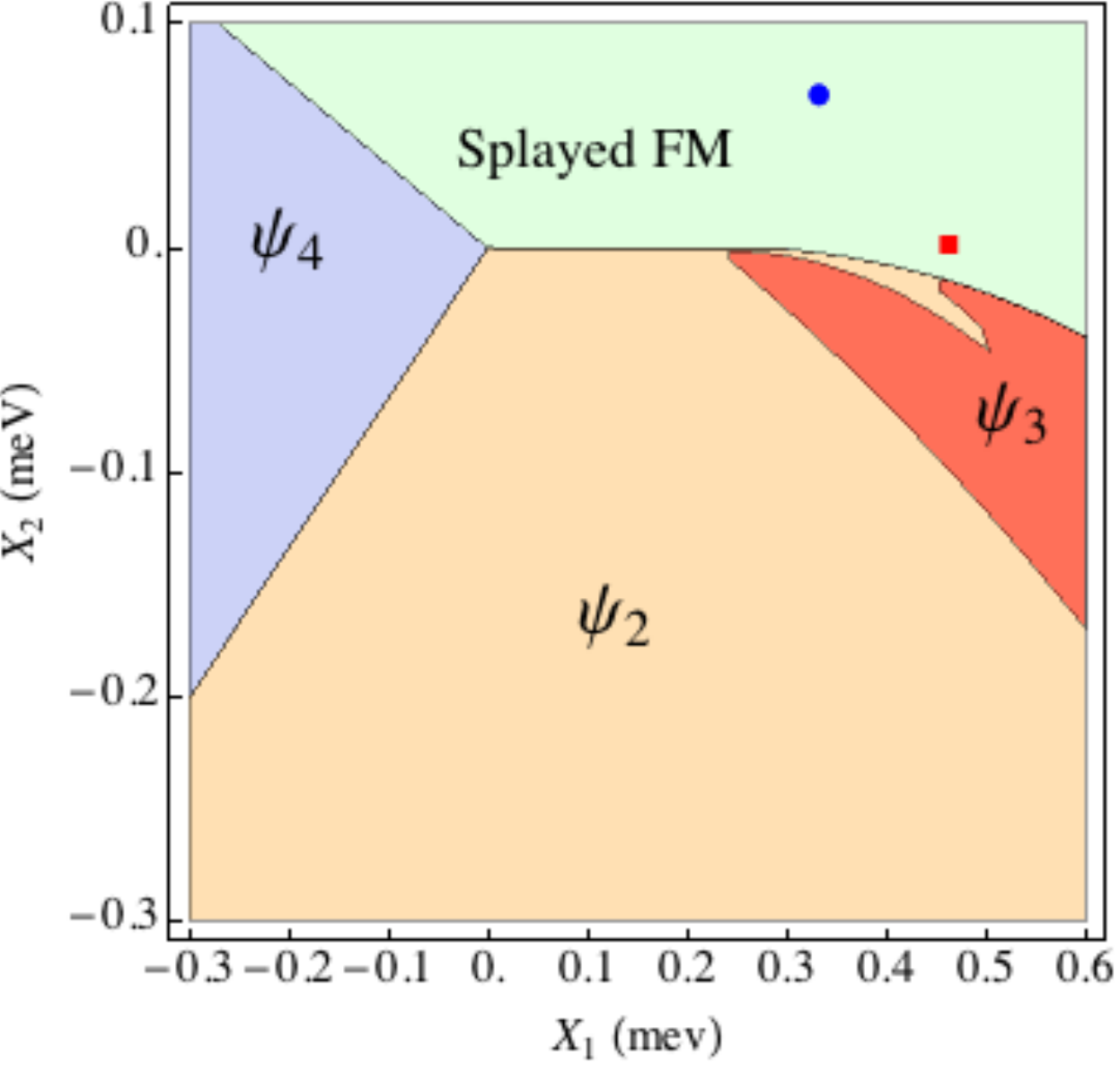}
\caption{Semiclassical phase diagram in the space of anisotropic
exchange parameters given in Eq.~(\ref{eq:Ham}). The origin of the
phase diagram is a Coulomb phase at the classical level. Three
long-ranged ordered phases appear in its vicinity: a splayed
(canted) ferromagnet (FM) with spontaneous polarization along one
of the cubic directions, the $\psi_4$(Palmer-Chalker)
\cite{McClarty2009} antiferromagnetic state and a ground state
with an accidental degeneracy in the $\Gamma_5$ manifold, resolved
by quantum fluctuations into long-range ordered antiferromagnetic
states $\psi_2$ (as in the ordered phase of Er$_2$Ti$_2$O$_7$) or
$\psi_3$, both with only a discrete degeneracy
\cite{McClarty2009}. All these phases have ordering wavevector
$\bm{q}=\bm{0}$. Two points are marked on the phase diagram
corresponding to the exchange values proposed by Ross {\it et al.}
\cite{Ross2011a} (blue circle) and those obtained here (red
square). The $X_1$ and $X_2$ coordinates are linear combinations
of the primary couplings chosen such that the phase diagram
includes both sets of Hamiltonian parameters and also the Coulomb
phase at the origin. In particular, $X_1=\left\{-0.9198, -0.3834,
-0.084, 0 \right\}$ and $X_2=\left\{0.349, -0.7233, -0.5204,
0.2904\right\}$ in the space of couplings $\left\{J_{zz},
J_{\pm\pm}, J_{\pm},J_{z\pm} \right\}$.} \label{fig:PhaseDiagram}
\end{figure}

\subsection{Mean-Field Phase Diagram}

The fit to the exchange parameters fixes the semiclassical ground
state of Yb$_2$Ti$_2$O$_7$ to be a canted ferromagnet with an
ordering wavevector $\bm{q}=\bm{0}$ and spontaneously chosen net
polarization along one of the cubic axes. The moments are
non-collinear, tilted towards the the local [111] axes. The mean
field $T_c \approx 2.95$~K is far in excess of the observed
transition temperature in the material. We have computed the
ground state phase diagram for the exchange model
Eq.~(\ref{eq:Ham}) in the vicinity of these exchange parameters in
order to point out proximate phases. To this purpose, we choose a
highly symmetric point in the space of anisotropic couplings that
harbors a Coulomb phase at the classical level corresponding to
couplings $J_{zz}=1$, $J_{\pm\pm}=0.5$, $J_{\pm}=0.25$ and
$J_{z\pm}=-1/\sqrt{8}$ \cite{Benton2016}. We rescale these
couplings so that $J_{z\pm}$ matches the value extracted from the
experimental data namely $-0.162$ meV. Then we choose to plot the
phase diagram in Fig.~\ref{fig:PhaseDiagram} in a plane through
the space of couplings containing both the exchange values
proposed by Ross {\it et al.} \cite{Ross2011a} (blue circle) and
those obtained here (red square). Evidently, both sets of
parameters place Yb$_2$Ti$_2$O$_7$ in the same phase, which is the
same as the one determined experimentally. The principal
difference between the two sets of exchange parameters is that the
set determined in this work lies very close to the phase boundary
with the $\psi_3$ state with antiferromagnetic order. While
apparent from the phase diagram, one may confirm that the
difference between the mean field energy of the ground state and
the energy of the $\psi_3$ state is smaller for the exchange
parameters determined in this work.
%%%%%%%%%%%%%%%%%%%%%%%%%%%%%%%%%%%%

%%%%%%%%%%%%%%%%%%%%%%%%%%%%%%%%%%%%%%%%%%%%%%%%%%%%%%%%%%%%%%%%%%%%%%%%
\subsection{Curie-Weiss Temperature}

The Curie-Weiss temperature obtained from a high temperature
expansion for the Hamiltonian in (\ref{eq:Hamiltonian}) is
\cite{Ross2011a}
\begin{align*}
& 2k_{\rm B} (2g_{\perp}^2 + g^2_{\parallel}) \Theta_{\rm CW}
\\ & =  \left(  g_{\parallel}^2 J_{\rm{zz}}  - 4g_{\perp}^2
\left( J_{\pm} + 2J_{\pm\pm} \right)  - 8\sqrt{2} g_{\perp}
g_{\parallel} J_{z\pm}    \right).
\end{align*}
For the parameters in Ref.~\cite{Ross2011a}, $\Theta_{\rm
CW}=312$~mK whereas we find $641$~mK for the parameters extracted
in this study. This calculated value is to be compared with
experimental values of $400$~mK \cite{Blote} and $750$~mK
\cite{Hodges2001}.
%%%%%%%%%%%%%%%%%%%%%%%%%%%%%%%%%%%%%%%%%%%%%%%%%%%%%%%%%%%%%%%%%%%%%%%%

\subsection{Magnon Decay Matrix Element}

The spin structure of Yb$_2$Ti$_2$O$_7$ is noncollinear for the
entire range of fields explored in the experiment reported here.
Then, in the local quantization frame there are terms for an
isotropic exchange Hamiltonian that couple $x$ and $z$ components
of the spins. As a consequence, the $1/S$ expansion in terms of
Holstein-Primakoff (HP) bosons contains cubic interaction terms
which take one magnon into two or {\it vice versa}. In our case,
such terms arise as a consequence of noncollinearity {\it and}
anisotropic exchange in the global frame.

\begin{figure*}[tb!]
\begin{center}
\includegraphics[width=\textwidth]{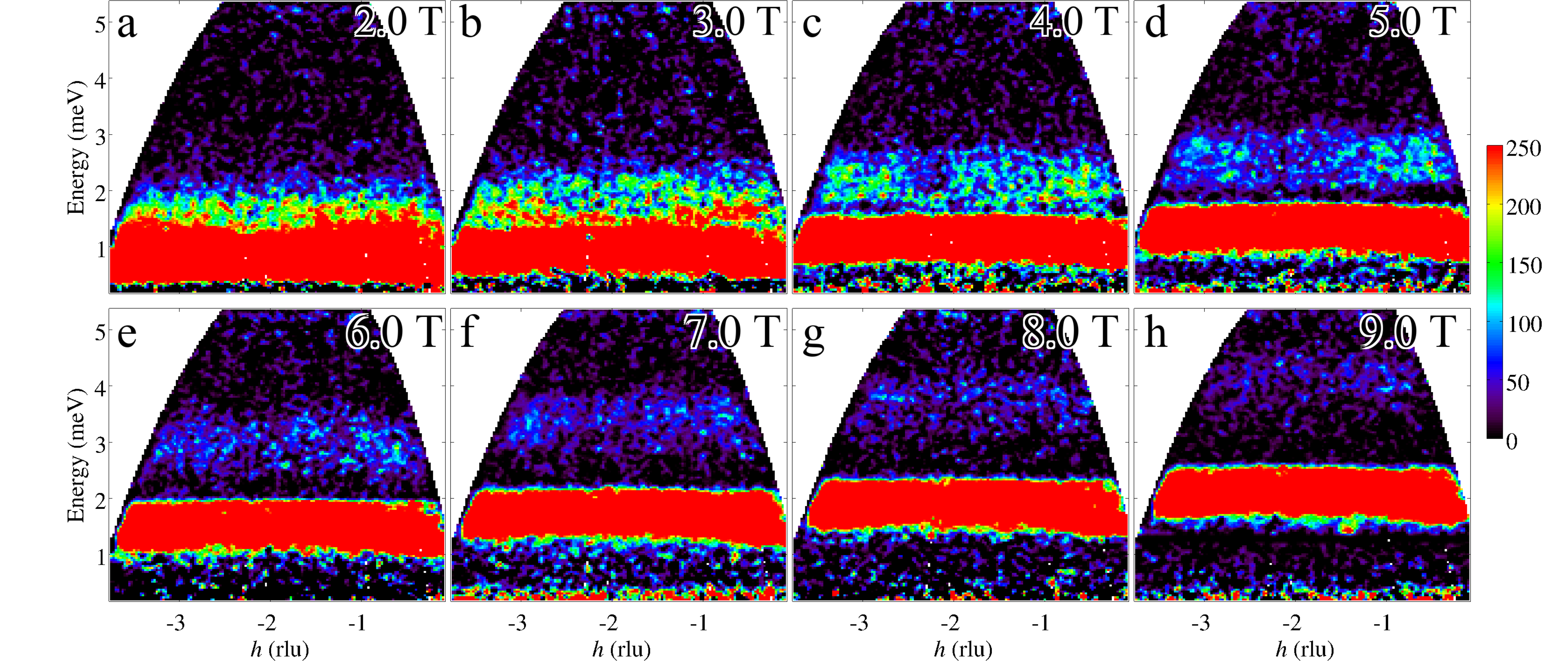}
\caption{INS data collected in a high-flux and low neutron
absorption configuration ($E_i=6.3$~meV) to maximize the
sensitivity to observe the weak two-magnon continuum scattering.
The strong (out-of-range) signal at intermediate energies is due
to one-magnon excitations, the weaker intensity band of scattering
at higher energies is due to two-magnon excitations. Note the
continuum scattering increases in intensity and moves to lower
energies upon decreasing field. An energy scan through the 7~T
data is shown in Fig.~\ref{fig:1}h).} \label{fig:Intensity6.3meV}
\end{center}
\end{figure*}

The spin wave dispersions, computed to quadratic order in HP
bosons, are renormalized by interaction terms. In the case of
cubic terms, there is a self-energy contribution
$\boldsymbol{\Sigma}\left( \bm{k}, \omega_{\bm{k}} \right)$ to the
spectrum coming from bubble diagrams. Schematically
\[  \boldsymbol{\Sigma}\left( \bm{k}, \omega_{\bm{k}}  \right)
\sim \sum_{\bm{q}}  \frac{ \vert \Gamma\left( \bm{k}, \bm{q}
\right)\vert^2 } { \omega_{\bm{k}}  -  \omega_{\bm{q}} -
\omega_{\bm{k}-\bm{q}} + i0^+   }   \] where $\Gamma$ is the
amplitude of the cubic vertex and for clarity we have omitted the
individual mode labels of the three magnon energies, which in
principle can each belong to a different dispersion mode $m$.
There are singularities in the integrand whenever the single
magnon energy $\omega_{\bm{k}}$ overlaps with the two-magnon
continuum $\omega_{\bm{q}} + \omega_{\bm{k}-\bm{q}}$. When this is
the case, magnon decay processes become kinematically allowed and
one expects a renormalization of the spectrum and also magnons
acquire a finite lifetime resulting in a broadening of the magnon
peaks at the nominal energy $\omega_{\bm{k}}$. The extent of these
effects depends strongly on the density of states of two magnon
decays. In Yb$_2$Ti$_2$O$_7$ the estimated threshold field below
which the two-magnon continuum overlaps with the highest-energy
one-magnon dispersion mode is $2.3$~T and below this field
experiments observe a substantial broadening of the highest energy
magnon lineshape in the overlap region (see Sec.~S7).

%%%%%%%%%%%%%%%%%%%%%%%%%%%%%%%%%%%%%%%%%%%%%%%%%%%%%%%%%%%%%%%%%%%%%%%%
\section{S6. Two-Magnon Scattering Continuum at High Fields}

\begin{figure}[tb!]
\begin{center}
\includegraphics[width=\columnwidth]{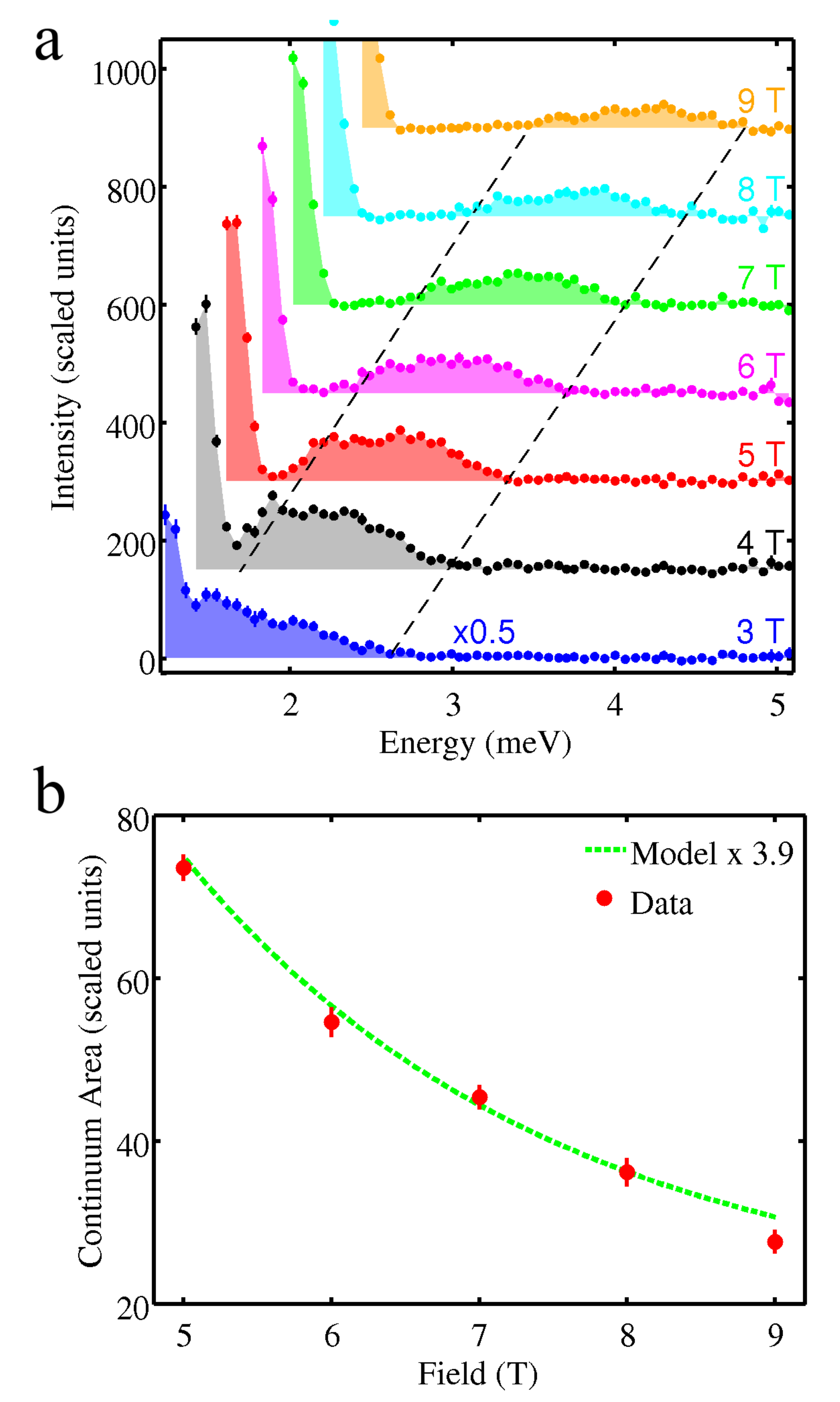}
\caption{(a) Energy scans through the intensity maps in
Fig.~\ref{fig:Intensity6.3meV} ($h=[-4,0]$) showing the two-magnon
scattering continuum increasing in intensity and being displaced
to lower energies upon decreasing field from top to bottom traces
(shading emphasizes the magnetic intensity). For clarity traces
are offset vertically (by +150) as a function of increasing field.
Dashed lines indicate the linear field dependence of the extracted
continuum boundaries. Intensities in the lower trace are
multiplied by 0.5 to fit on the same scale. b)
Experimentally-extracted energy integrated area of the continuum
scattering (from scans such as in Fig.~\ref{fig:1}h) as a function
of internal field and the corresponding spin-wave prediction for
two-magnon scattering (solid line) times an overall scaling factor
given in the legend.}\label{fig:ContinuumField}
\end{center}
\end{figure}

The INS intensity maps at high field showed a distinct continuum
intensity signal at energies corresponding to twice the one-magnon
energies, identified with neutrons scattering by creating a pair
of magnons. The intensity map in Fig.~\ref{fig:Intensity6.3meV}e)
shows this scattering contribution at 6~T, note the weak band of
scattering intensity centered around 3~meV, which shifts in energy
upon varying field (compare with data in panels at other fields).
Energy scans observing directly the shift in energy and intensity
increase upon lowering field are shown in
Fig.~\ref{fig:ContinuumField}a) (shaded areas).

\begin{figure}[tbp!]
\begin{center}
\includegraphics[width=\columnwidth]{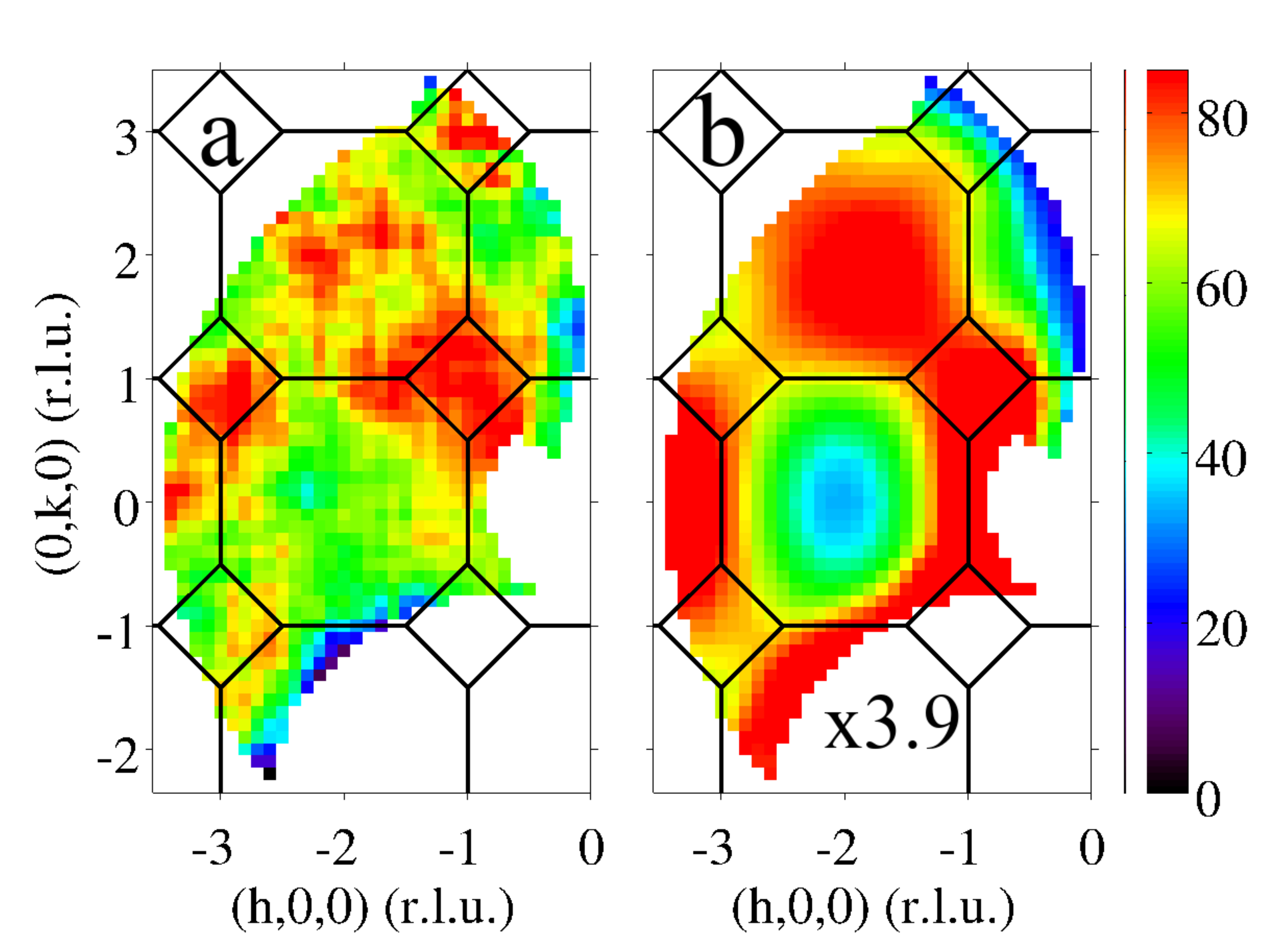}
\caption{(a) Wavevector-dependence of the continuum scattering
intensity in the $(hk0)$ plane at 5~T. Data comes from a Horace
scan averaged over $E=[2.03,3.06]$~meV and $l=[-0.2,0.2]$. b)
Corresponding intensity map of the two-magnon continuum scattering
in spin-wave theory, with the same overall scale factor as in
Fig.~\ref{fig:ContinuumField}. Solid lines show intersections with
the FCC Brillouin zone boundaries.} \label{fig:ContinuumSlice}
\end{center}
\end{figure}

A quantitative comparison with spin-wave theory for
non-interacting magnons in Fig.~\ref{fig:1}h)(solid line) shows
that the intensity of the continuum scattering relative to the
one-magnon intensity is underestimated, and also that the
continuum lineshape profile is different (more intensity at the
lower boundary), suggesting that inclusion of magnon-magnon
interactions may be required to account for those features. Apart
from an overall renormalization of the relative continuum
intensity its strong increase upon lowering field is well captured
by spin-wave theory (solid line) in
Fig.~\ref{fig:ContinuumField}b). The wavevector-dependence of the
continuum intensity at 5~T is shown in
Fig.~\ref{fig:ContinuumSlice}a), the intensity has a local minimum
near ($\bar{2}00$), and this feature is well captured by the
spin-wave prediction for the two-magnon intensity (panel b).
%\vspace{-2.5cm}

\section{S7. Magnon Decay and Dispersion Renormalization at
Intermediate Fields}

\begin{figure}[htbp!]
\begin{center}
\includegraphics[width=\columnwidth]{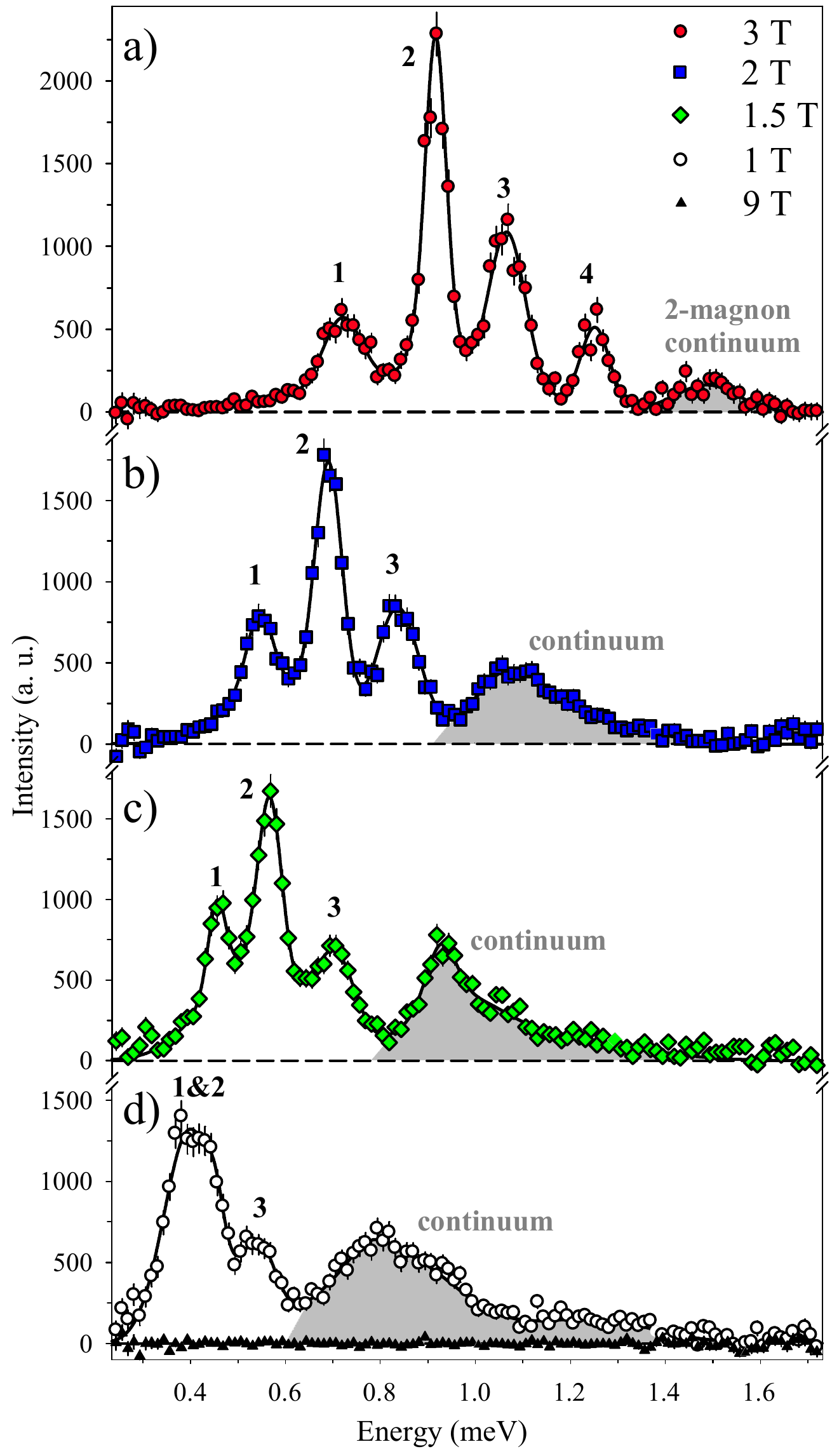}
\caption{Energy scan illustrating magnon decay and dispersion
renormalization effects below the field where one- and two-magnon
phase spaces overlap. The spacing between the lowest three sharp
modes (labelled 1-3) is significantly reduced as the whole group
shifts to lower energies upon reducing field (top to bottom). At
the same time the relative separation to the high-energy continuum
scattering (shaded area) reduces, at 2~T (panel b) the fourth
magnon mode has already ``dissolved'' in the continuum, which
rapidly grows in intensity upon lowering field. Data in all scans
comes from the intensity maps at the corresponding fields in
Fig.~\ref{fig:FieldSlices2.5} for $h=[-2.2,-1.5]$ and
$l=[-0.4,0.4]$. Filled triangles in bottom panel show 9~T data to
illustrate the quality of the non-magnetic background
subtraction.} \label{fig:EnergyScan_MagnonDecay}
\end{center}
\end{figure}

An overview of the field-dependence of the spin dynamics is
plotted in Fig.~\ref{fig:FieldSlices2.5} as a function of
increasing field in the top and every other subsequent row,
whereas the rows of panels immediately below show the spin-wave
calculation. Good agreement is found above $\sim$$3$~T, below this
field one- and two-magnon phase spaces overlap [see
Fig.~\ref{fig:1}i)] and more complex behavior occurs. The energy
scan in Fig.~\ref{fig:EnergyScan_MagnonDecay}a) at 3~T shows four
well-resolved sharp peaks followed by a weak scattering continuum
(shaded area) centered near 1.5~meV. Upon lowering field to 2~T
(panel b) the lower three peaks have shifted to lower energies,
whereas the fourth peak has merged with the continuum with a lower
boundary near 1~meV. We interpret this ``disappearance'' of the
highest-energy magnon mode as being due to its spontaneous decay
into two-magnon states. Upon lowering the field further to 1.5~T
(panel c) the whole pattern shifts to lower energies and
furthermore the energy spacing between the lower three peaks is
clearly smaller than at 3~T (panel a). At 1~T (d) the peaks 1 and
2 have almost merged and the spacing 1-3 is reduced further. This
magnon bandwidth narrowing is not captured by the (linear)
spin-wave approach, which predicts almost field-independent
bandwidths, compare Fig.~\ref{fig:FieldSlices2.5}c-d) with g-h).

We attribute the dispersion renormalization to the increase of
quantum fluctuations upon lowering field, also manifested in the
continuum intensity now becoming comparable to that of the sharp
modes. Physically, a reduction in the magnon bandwidth could be
interpreted as a reduction in the kinetic energy that
one-spin-flip states gain from coherently hopping across the
lattice sites, or an effective inhibition of such coherent
propagation due to the increased quantum fluctuations at low
field. Given the close proximity between the sharp modes and the
continuum boundary at those low fields it may be possible that the
dispersion renormalization effects observed might be captured, at
least partly, by including interactions between the one-magnon
states and the higher-energy continuum scattering.

\subsection{Scattering Continuum in Zero Field}

The discrepancy between the linear spin-wave prediction and the
data becomes even more dramatic in the region of very low fields.
In zero field the spin-wave model predicts sharp modes in the
range 0.21-0.61~meV, whereas the data shows a broad scattering
continuum throughout this range with considerable scattering
weight at lower energies and also extending up to 1.5~meV, compare
Fig.~\ref{fig:FieldSlices2.5}a and e). The continuum scattering
lineshapes are clearly apparent in Fig.~\ref{fig:1}e) (red
symbols) with no clear sharp modes seen in this energy range, in
clear contrast with the spin-wave prediction.

\begin{figure*}[tbp!]
\begin{center}
\includegraphics[width=\textwidth]{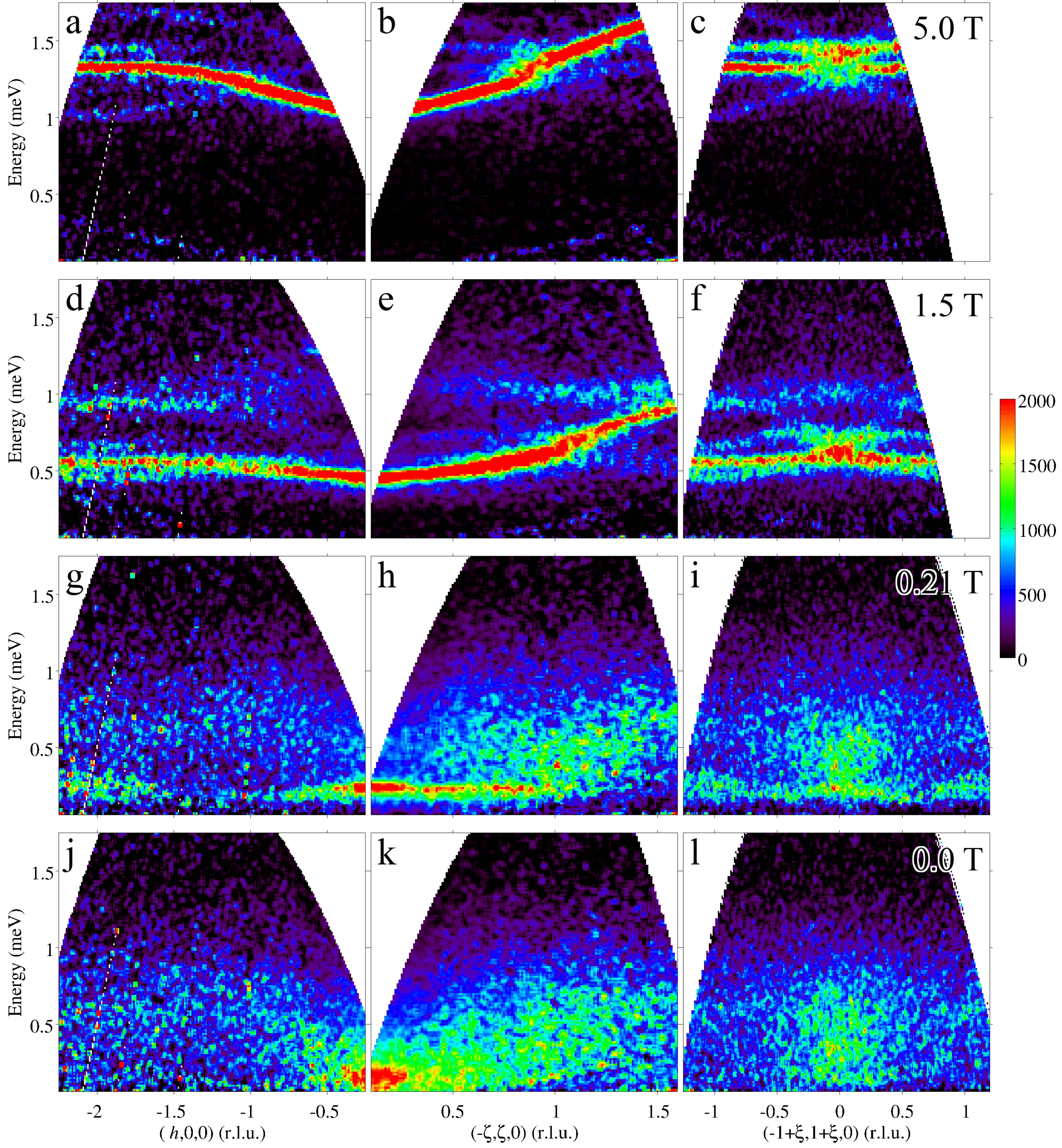}
\caption{Field-dependence of the excitation spectrum for several
high-symmetry directions in the ($hk0$) plane, field decreases
from top to bottom. Corresponding spin-wave calculations are shown
in Fig.~\ref{fig:horace_overview_LSWT}. Data was extracted from a
Horace scan with $E_i=2.5$~meV. Intensities are averaged for
transverse wavevectors in the range $l,k=[-0.2,0.2]$ for the
($h00$) direction, and similar ranges for the other directions.}
\label{fig:horace_overview}
\end{center}
\end{figure*}

\begin{figure*}[tbp!]
\begin{center}
\includegraphics[width=\textwidth]{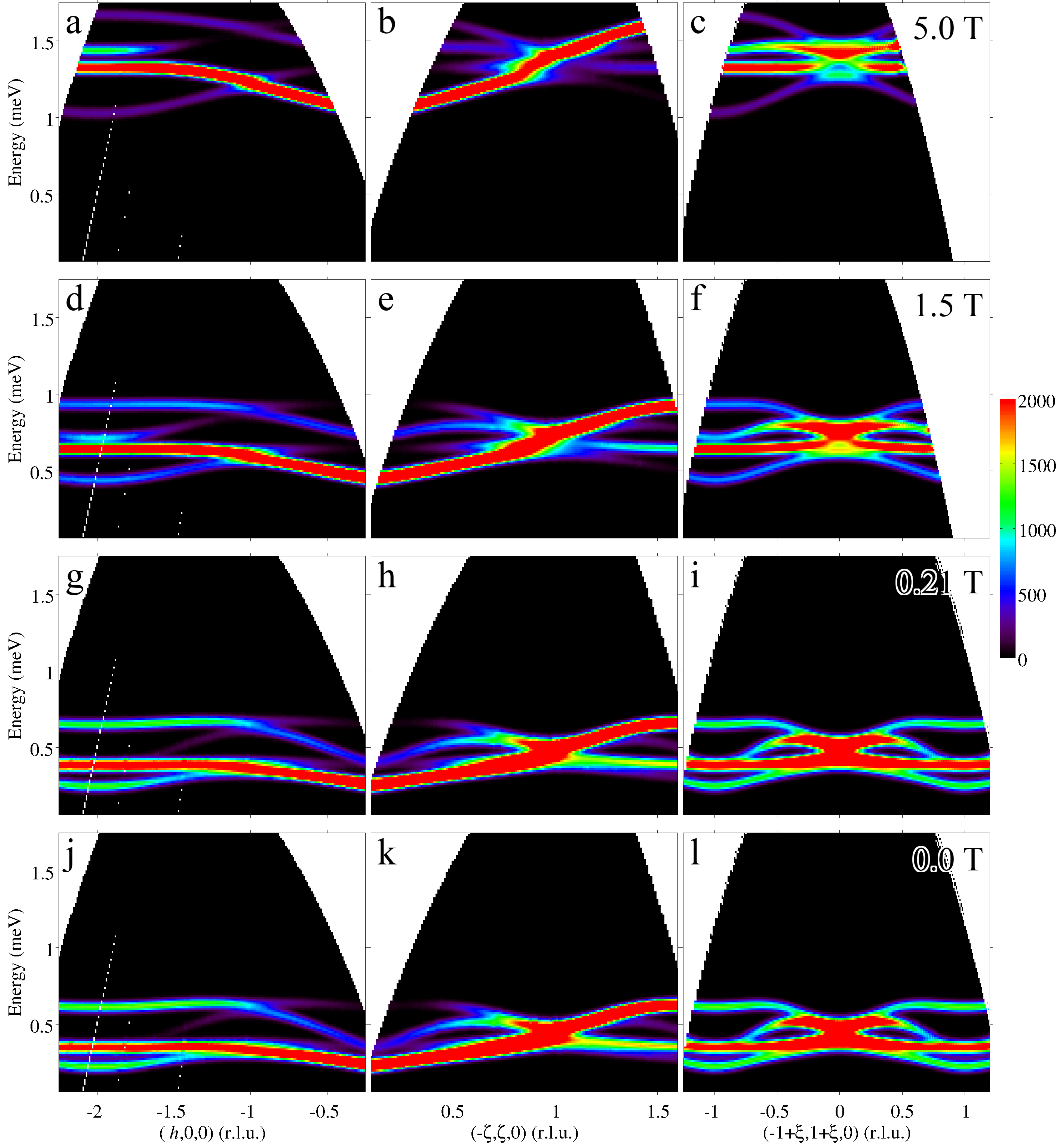}
\caption{Spin-wave calculations corresponding to the data plotted
in Fig.~\ref{fig:horace_overview}. Calculations are performed for
demagnetization-corrected fields and include the magnetic form
factor and convolution with an estimated energy resolution.}
\label{fig:horace_overview_LSWT}
\end{center}
\end{figure*}

\begin{figure*}[th!]
\begin{center}
\includegraphics[width=\textwidth]{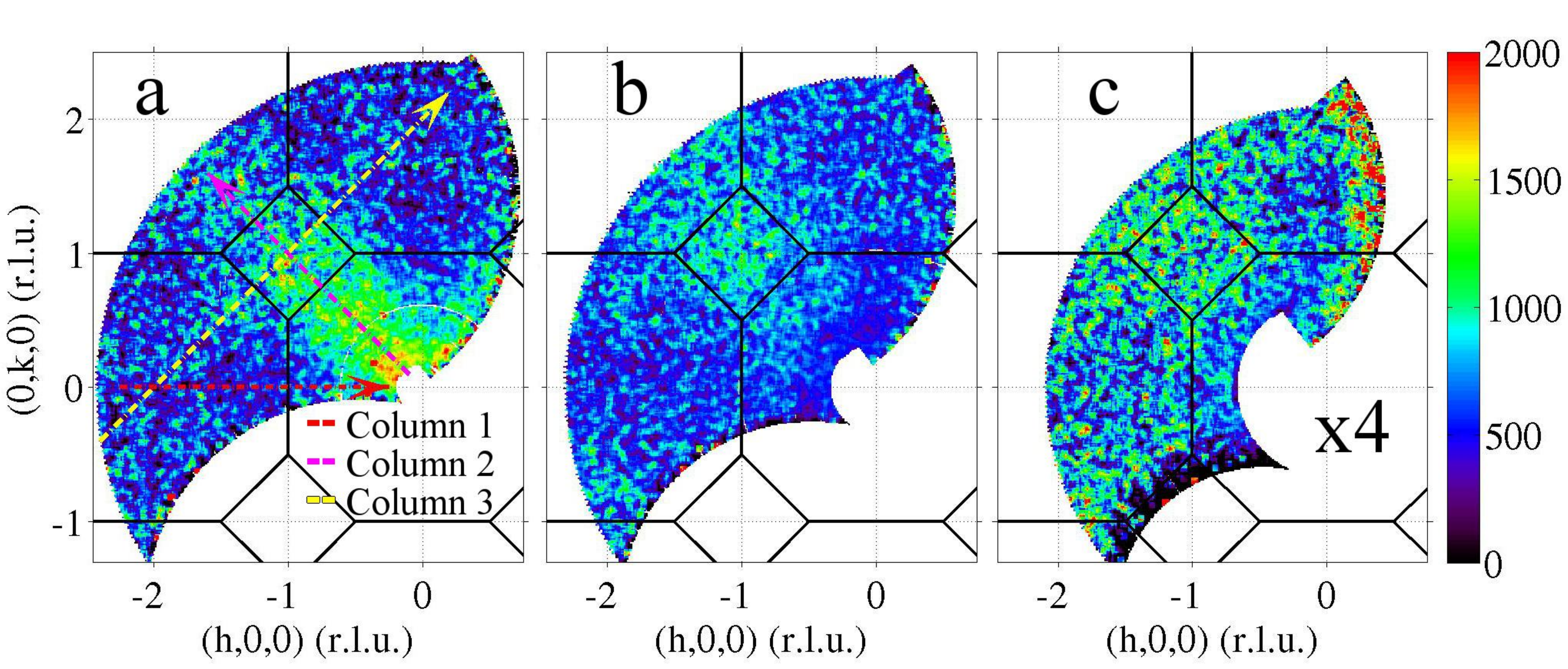}
\caption{Wavevector-dependence of the zero field continuum
scattering intensity in the ($hk0$) plane at a) low
($[0.04,0.4]$~meV), b) intermediate ($[0.4,0.8]$~meV) and c) high
energies ($[0.8,1.5]$~meV). Intensities are averaged over
transverse wavevectors in the range $l=[-0.2,0.2]$. In panel c)
intensities are scaled $\times 4$ to bring them on the same range
as the other panels. Solid black lines show intersection with FCC
Brillouin zone boundaries. Dashed arrowed lines in a) labelled 1-3
show directions along which the spectrum is plotted in
Fig.~\ref{fig:horace_overview} columns 1-3.}
\label{fig:continuum_horace_slices}
\end{center}
\end{figure*}

An overview of how the spectrum evolves as a function of field for
several wavevector directions in the ($hk0$) plane extracted from
a Horace data volume is shown in Fig.~\ref{fig:horace_overview}.
Notice the contrast between the 5~T data (top row) dominated by
well-defined sharp modes with a large gap, and zero field (bottom
row) where an extended scattering continuum dominates. At a
relatively small applied field of 0.21~T a clear sharp mode is
stabilized near 0.22~meV for small $|\bm{Q}|$ see
Fig.~\ref{fig:horace_overview}h) (bottom left), but the continuum
scattering up to 1.5~meV is essentially unchanged compared to zero
field (compare with panel k). Continuum scattering is observed at
all wavevectors probed, with some clear intensity modulations
illustrated in Fig.~\ref{fig:continuum_horace_slices}, the
intensity is strongest near the ($hh0$) line for energies below
$\sim$0.4~meV (panel a), at higher energies the intensity appears
more uniformly distributed (panels b-c). The zero and low field
spin dynamics is in sharp contrast with linear spin-wave theory,
which would predict a spectrum dominated by dispersive sharp
magnon modes, compare Figs.~\ref{fig:horace_overview} and
~\ref{fig:horace_overview_LSWT} (bottom rows). We interpret this
dramatic quasiparticle breakdown over a large part of the
Brillouin zone as an indication that fluctuations become very
strong in the low field regime, presumably due to the dominant
``quantum'' exchange term $J_{z\pm}$, with the consequence that a
semiclassical linear spin-wave description becomes inadequate to
capture the spin dynamics.

%%%%%%%%%%%%%%%%%%%%%%%%%%%%%%%%%%%%%%%%%%%%%%%%%%%%%%%%%%%%%%%%%%%%%%%%
\section{S8. Heat Capacity Measurements}
\label{heatcapacity}

Heat capacity measurements were collected using a home-made setup
based on the AC technique \cite{Rost,Sullivan1968}. The sample was
a flat-plate 9.7~mg single crystal of Yb$_2$Ti$_2$O$_7$ with
approximate dimensions $1.85\times1.61\times0.45$~mm$^3$ cut from
the same piece as the crystal used for the INS measurements.
Cooling was provided by an Oxford Instruments Kelvinox25 dilution
refrigerator, of base
temperature $\sim$$30$~mK, equipped with a $16$~T superconducting
magnet, with the sample aligned such that the magnetic field was
applied perpendicular to the sample plate, along the [001] crystal
direction.

The heat capacity setup contained a Lakeshore RX-$102$A-BR
Ruthenium Oxide thermometer mounted on the sample using Apiezon N
grease, which was then connected to a $5\times5\times0.150$~mm$^3$
$99.95\%$ pure Ag platform. A $120~\Omega$ strain gauge was
connected to the bottom of the platform using GE $7031$ varnish
and used as a heater to apply an oscillating temperature at a
frequency of $5\times10^{-3}$~Hz. A $1$~cm $99.99\%$ pure Pt wire
was used to establish a heat link between the platform and an OFHC
Copper heat sink connected to the mixing chamber of the dilution
fridge. The thermometer was calibrated down to $30$~mK in zero
field against a calibrated Lakeshore RX-$102$B-CB thermometer.
Field calibrations of the thermometer were performed using
constant temperature magnetic field sweeps. The measured heat
capacity was normalized into absolute units by calibration against
the known specific heat of a standard sample measured in the same
setup. The applied magnetic field values $\mu_0H_{\rm app}$ were
corrected for demagnetization effects to obtain the net internal
fields $\mu_0H_{\rm int}$ as discussed in Sec.~S9.

\begin{figure}[h!]
\begin{center}
\includegraphics[width=\columnwidth]{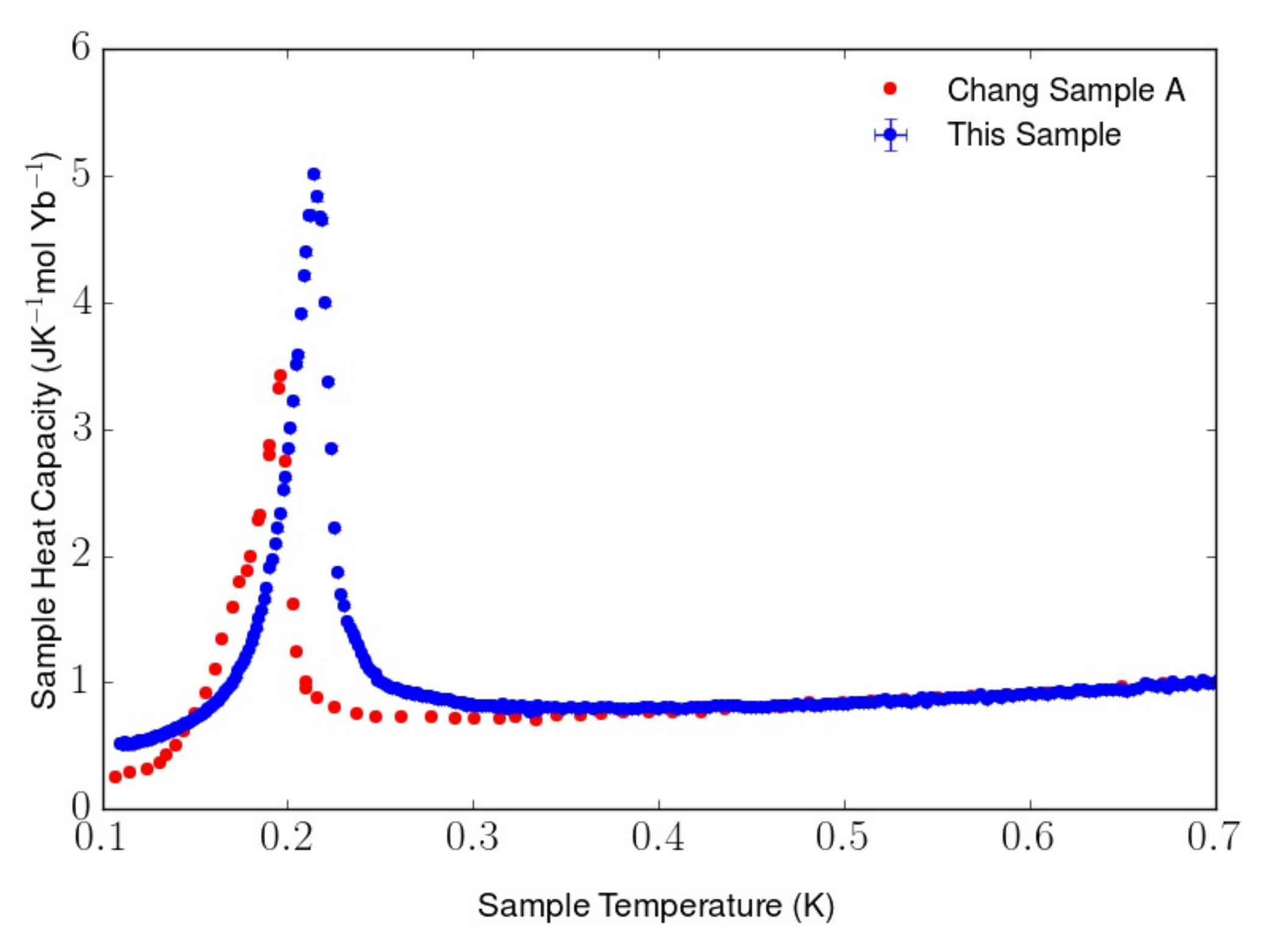}
\caption{Specific heat as a function of temperature in zero
magnetic field (blue symbols) compared to earlier reports
\cite{Chang2012} (red symbols).} \label{fig:HC1}
\end{center}
\end{figure}

The heat capacity was measured at fixed applied field as a
function of increasing temperature. Fig.~\ref{fig:HC1} displays
the obtained heat capacity in zero field (blue symbols). The
behavior is comparable to that reported in Ref.~\cite{Chang2012}
(red symbols) on a crystal where the sharp low-temperature anomaly
has been identified with the onset of spontaneous canted
ferromagnetic order. Systematic studies of samples of various
purities have shown that more stoichiometric
samples\cite{Ross2014,Arpino2017} display a single sharp peak in
the heat capacity (at temperatures up to 0.26~K), whereas samples
believed to be affected by substantial structural
disorder/stuffing/oxygen non-stoichiometry show rather different
behavior, with only a broad peak or multiple
peaks\cite{Ross2012a}. The presence of a single sharp peak in the
heat capacity of our sample is indicative of a sharp transition to
a well-developed magnetic order, suggesting that structural
disorder effects are rather small and the magnetic behavior is
representative of the high-purity limit.

Fig.~\ref{fig:2} shows heat capacity measurements in applied
magnetic field. Above $0.1$~T the sharp peak observed in zero
field is completely suppressed, replaced by a broad Schottky
feature \cite{Gopal1966}. Upon increasing field the
low-temperature tail of the specific heat is progressively
suppressed and the Schottky anomaly moves to higher temperatures,
both are indications of a gap in the excitation spectrum, which
increases upon increasing field. To capture this trend we compare
the data to the behavior expected for a system with an excited
level at energy $\Delta$ above the ground state,
\begin{equation}
C(T)=R \left( \frac{\Delta}{k_{\rm B}T}\right)^2
\frac{e^{-\Delta/k_{\rm B}T}}{\left( 1+e^{-\Delta/k_BT}\right)^2}.
\label{eq:two-level}
\end{equation}
\begin{figure}[tb!]
\begin{center}
\includegraphics[width=\columnwidth]{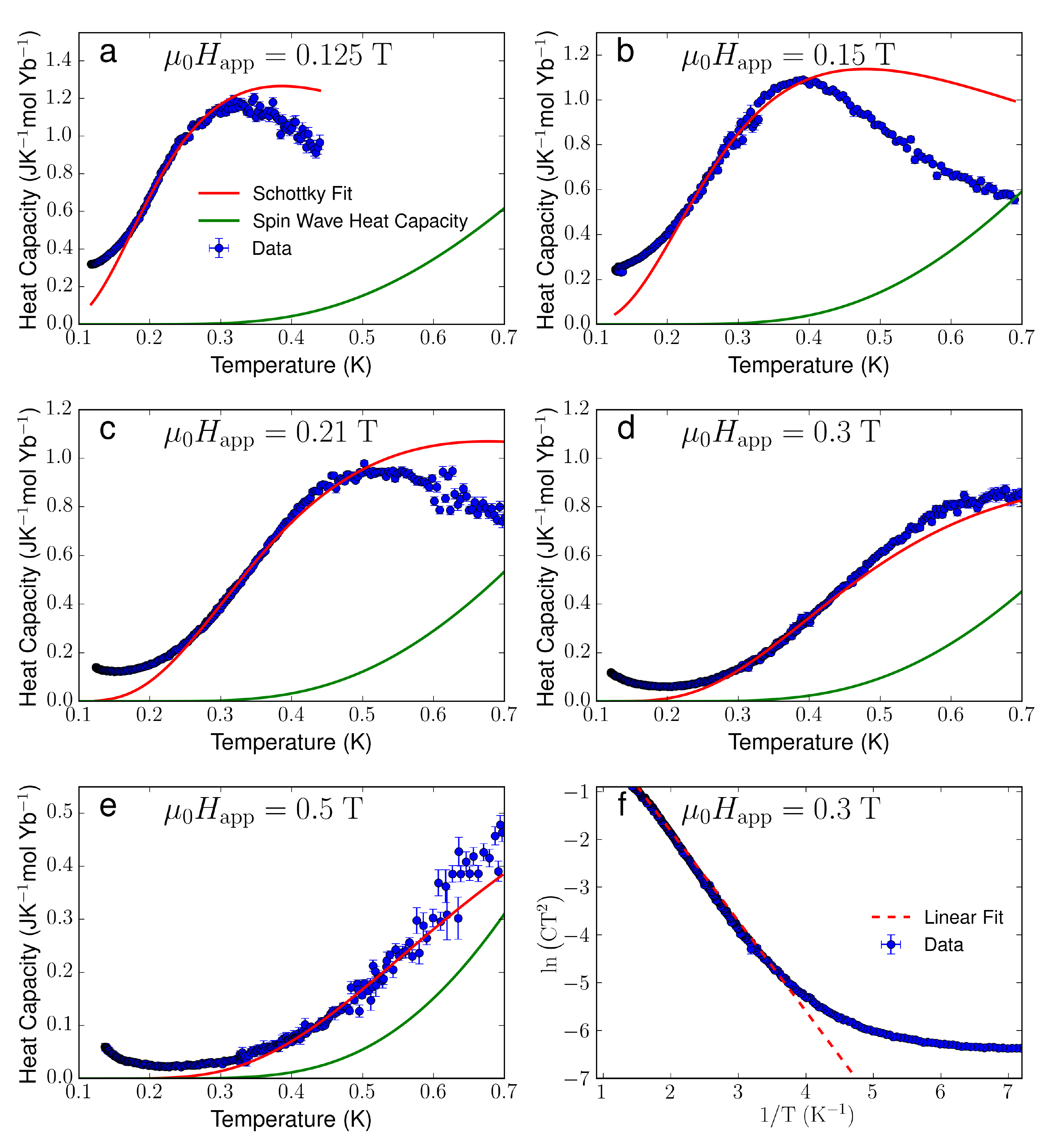}
\caption{a-e) Specific heat as a function of temperature and
magnetic field $\parallel [001]$. Solid lines show fits of the
rising part of $C(T)$ to a two-level system,
eq.~(\ref{eq:two-level}), solid green lines show the spin-wave
prediction. Data points are the raw specific heat minus an
estimate of the non-magnetic contribution obtained from
measurements at high field (lower trace in Fig.~\ref{fig:2}). f)
Same data as in d) plotted now as $\ln\left(CT^2\right)$ vs. $1/T$
to expose the near-linear dependence (dashed line) predicted by
(\ref{eq:two-level}) for $T\ll \Delta/k_{\rm B}$.}
\label{fig:fitsHC}
\end{center}
\end{figure}
%%%%%%%%%%%%%%%%%%%%%%%%%%%%%%%%%%%%%%%%%%%%%%%%%%%%%%%%%%%%%%%%%%%%%%%%
This form was fitted to the measured $C(T,B)$ data in the
temperature region up to the broad peak maximum and avoiding the
very low-temperature part, where quadrupolar contributions lead to
a $1/T^2$ behavior, seen already in earlier measurements
\cite{Blote}. The non-magnetic contribution was estimated from the
measured $C(T)$ at high field (lowest trace in Fig.~\ref{fig:2})
when the spin gap is $\sim$0.4~meV, such that the population of
thermally excited magnon states over the whole temperature range
of the heat capacity measurements (up to $0.7$~K) is negligible.
The fits are illustrated in Fig.~\ref{fig:fitsHC}
\begin{figure}[tb!]
\begin{center}
\includegraphics[width=\columnwidth]{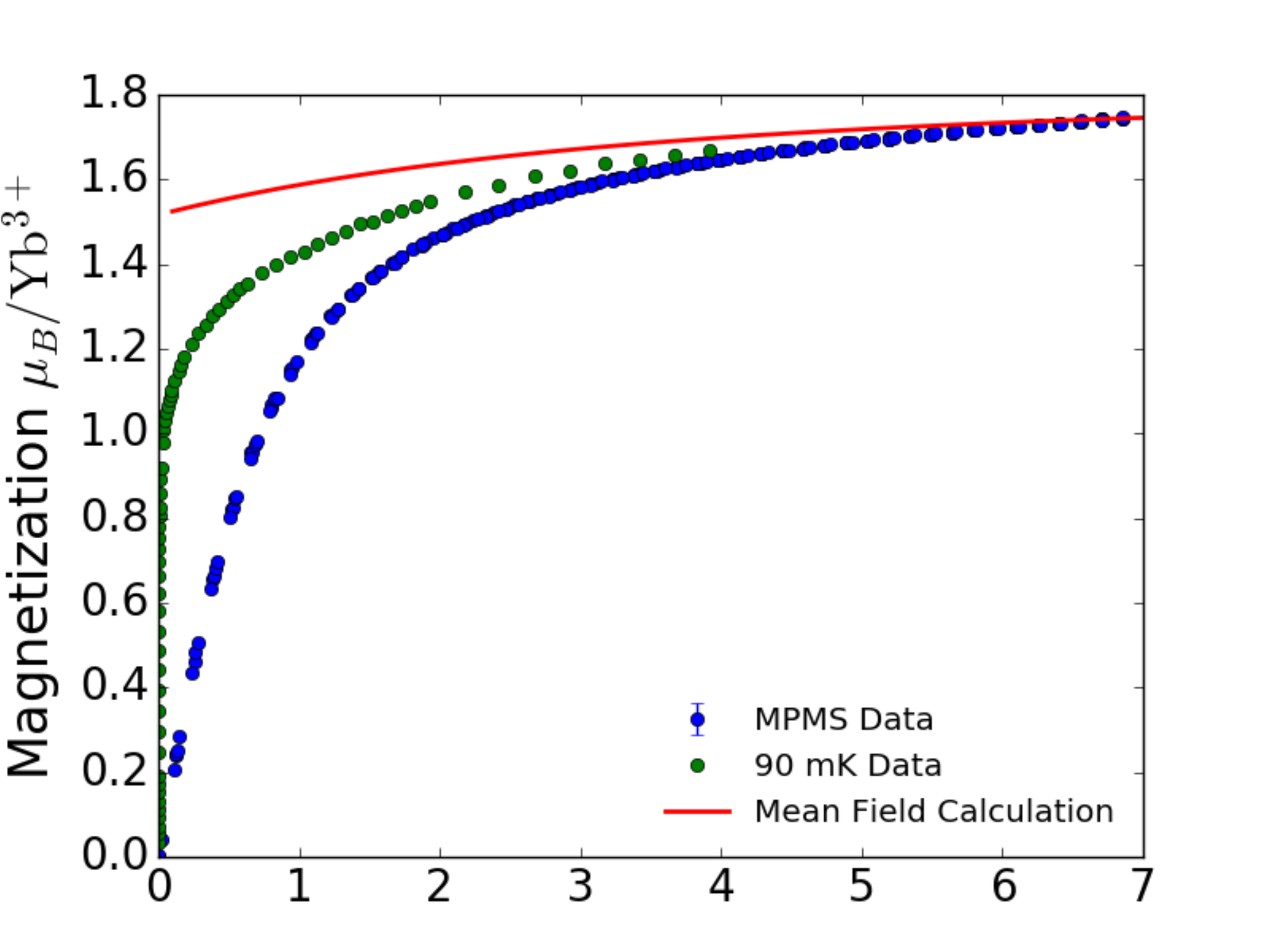}
\caption{Magnetization as a function of magnetic field along [001]
(blue points, 1.8~K). Green points are lower temperature (0.09~K)
data from Ref.~\cite{Lhotel2014}. The red solid line is the
zero-temperature mean-field calculation, which reproduces well the
observed magnetization value near saturation.
\label{fig:Magnetization}}
\end{center}
\end{figure}
and give a good parameterization of the rising part of $C(T)$. The
fitted pre-factor $R$ is systematically lower than the expected
molar gas constant, a reduction would be expected as INS
measurements [see Fig.~\ref{fig:1}b)] show a density of states
that is not concentrated solely in a single level at the gap
energy $\Delta$ as assumed by (\ref{eq:two-level}), but is in fact
extended over a wide energy range above the gap. The gap extracted
from the heat capacity data is plotted in Fig.~\ref{fig:2}(inset)
and shows a monotonic increase in field.

\begin{table*}[htbp!]
\begin{center}
\caption{Internal field as a function of applied field for the
heat capacity measurements.} \label{tab:demag_hc}
\begin{tabular}{|c|c|c|c|c|c|c|c|c|c|c|c|c|c|}
\hline
$\mu_0H_{\rm app}$(T)&$0.05$&$0.075$&$0.1$&$0.125$&$0.15$&$0.21$&$0.30$&$0.5$&$1.5$\\
\hline
$\mu_0H_{\rm int}$ (T)&$<0.001$&$0.0013(2)$&$0.007(1)$&$0.019(2)$&$0.035(3)$&$0.084(3)$&$0.165(4)$&$0.354(4)$&$1.330(5)$\\
\hline
\end{tabular}
\end{center}
\end{table*}
\begin{table*}[htb!]
\begin{center}
\caption{Internal field (typical uncertainty $\sim$0.002~T) as a
function of applied field for the INS measurements.}
\label{tab:demag_ins}
\begin{tabular}{|c|c|c|c|c|c|c|c|c|c|c|c|c|c|}
\hline
$\mu_0H_{\rm app}$ (T)&$0.21$&$0.5$&$0.75$&$1$&$1.5$&$2$&$3$&$4$&$5$&$6$&$7$&$8$&$9$\\
\hline
$\mu_0H_{\rm int}$ (T)&$0.166$&$0.45$&$0.698$&$0.946$&$1.44$&$1.94$&$2.94$&$3.94$&$4.93$&$5.93$&$6.93$&$7.93$&$8.93$\\
\hline
\end{tabular}
\end{center}
\end{table*}

As already shown by the INS data the low-field behavior of the
excitations cannot be captured by a spin-wave approach, a
scattering continuum dominates instead of sharp modes and moreover
there is a large density of states at energies much below the
predicted spin-wave gap in zero field of 0.2~meV, compare
Fig.~\ref{fig:FieldSlices2.5}a) with f). The failure of linear
spin-wave theory to capture the low-field behavior is also
dramatically illustrated by the specific heat data.
Fig.~\ref{fig:fitsHC}a-e) shows that the calculated spin-wave heat
capacity $C_{\rm SW}$ (solid green curves) significantly
underestimates the low-temperature heat capacity at low fields, as
expected if excited states existed below the predicted spin-wave
gap. The comparison also shows that upon increasing field the
spin-wave calculation becomes progressively closer to the data,
this is expected as in the limit of high enough fields where all
the low-energy excitations are well-defined, sharp magnons, then
we know from comparison to the INS data (at 5~T) that spin-wave
theory provides a very good description of the low-energy states,
compare Fig.~\ref{fig:1}d) and g). The comparison in
Fig.~\ref{fig:fitsHC}a-e) is consistent with the expectation that
$C_{\rm SW}$ approaches the measured $C(T)$ at high enough fields.

%%%%%%%%%%%%%%%%%%%%%%%%%%%%%%%%%%%%%%%%%%%%%%%%%%%%%%%%%%%%%%%%%%%%%%%%
\section{S9. Magnetization Measurements}
\label{magnetization} Magnetization data was collected in order to
provide further constraints on the overall scale of the
$g$-tensor. Measurements were performed at 1.8~K using a SQUID
magnetometer (Quantum Design MPMS) on a near-rectangular
27.50(2)~mg single crystal of Yb$_2$Ti$_2$O$_7$ with approximate
dimensions $2.09\times1.81\times1.01$~mm$^3$ cut from the same
crystal piece used for the heat capacity and INS experiments. The
sample was aligned such that the magnetic field was applied normal
to the largest sample face, along the $[001]$ crystallographic
axis. Fig.~\ref{fig:Magnetization} shows the obtained
magnetization curve, which gives a magnetic moment at the highest
field probed of $\mu_0H_{\rm int}=6.86$~T of $1.745(4)\mu_{B}$ per
Yb$^{3+}$ ion. Field values were corrected for demagnetization
effects as discussed below.

\subsection{Demagnetization corrections}
The applied magnetic fields in the magnetization, specific heat
and neutron scattering measurements were corrected for
demagnetization effects to obtain the internal fields $H_{\rm
int}=H_{\rm app}-NM$ where $N$ is the demagnetization factor and
$M$ is the magnetization volume density. For the heat capacity and
magnetization samples $N$ was calculated as $0.64$ and 0.49,
respectively, using analytical results for a rectangular prism
\cite{demag_prism}. The sample used for the INS measurements was a
cylinder with the magnetic field applied at an angle
$\theta\simeq32^\circ$ to the cylinder axis. Here the internal
field was approximated considering only the projection of the
demagnetization field along the applied field axis, giving an
effective $N \simeq
N_{\parallel}\cos^2\theta+N_{\perp}\sin^2\theta$, where the
demagnetization factors for the directions along and transverse to
the cylinder axis were calculated as $N_{\parallel}=0.13$ and
$N_{\perp}=0.44$, respectively, using analytical results for a
cylinder \cite{demag_cylinder}. The estimated internal fields for
the heat capacity and neutron data are listed in
Tables~\ref{tab:demag_hc} and ~\ref{tab:demag_ins}, where we used
as the reference magnetization curve $M$ vs. $\mu_0H_{\rm int}$
the reported data at 90~mK up to 4~T $\parallel[001]$ from
\cite{Lhotel2014} supplemented with our own magnetization data
extended up to 7~T at 1.8~K [see Fig.~\ref{fig:Magnetization}].
The quoted errors in the tables include an uncertainty in matching
the absolute scales of the above two magnetization curves such
that the estimated extrapolation of the 90~mK data to high fields
overlaps with the 1.8~K data above 5~T. Since the neutron
scattering data was collected at a slightly higher temperature of
150 mK where the magnetization is expected to be somewhat reduced
compared to 90 mK in the limit of low fields, in the region of low
fields the demagnetization corrections are therefore
overestimated, so the quoted internal fields in
Table~\ref{tab:demag_ins} are to be interpreted as a lower bound
and to become more accurate at high fields where the magnetization
is less sensitive to temperature. Similarly, for the heat capacity
temperature scans at constant applied field in Fig.~\ref{fig:2},
the internal fields quoted in Table~\ref{tab:demag_hc} are to be
interpreted as the values at the lowest temperatures at the start
of the scans, the actual internal fields are expected to increase
towards the applied field value upon increasing temperature (as
the magnetization decreases, so demagnetization corrections
reduce).

%%%%%%%%%%%%%%%%%%%%%%%%%%%%%%%%%%%%%%%%%%%%%%%%%%%%%%%%%%%%%%%%%%%%%%%%
%\bibliography{library}

\end{document}